\documentclass[10pt, preprint]{emulateapj}		
\usepackage{color}

\usepackage{style_wei_symbols}
\usepackage{style_wei}

\usepackage{natbib}
\begin{document}

\title{{Plasmoid Ejections} and Loop Contractions in an Eruptive M7.7 Solar Flare:
Evidence of Particle Acceleration and Heating in Magnetic Reconnection Outflows}

\author{Wei Liu\altaffilmark{1}$^,$\altaffilmark{2}, Qingrong Chen\altaffilmark{3}, Vah\'{e} Petrosian\altaffilmark{3}
   }	

\altaffiltext{1}{Lockheed Martin Solar and Astrophysics Laboratory,
  Building 252, 3251 Hanover Street, Palo Alto, CA 94304}
\altaffiltext{2}{W.~W.~Hansen Experimental Physics Laboratory, Stanford University, Stanford, CA 94305}
\altaffiltext{3}{Department of Physics, Stanford University, Stanford, CA 94305}

\shorttitle{Particle Acceleration in Reconnection Outflows}
\shortauthors{Liu et al.}
\slugcomment{Accepted by ApJ, March 04, 2013}

\begin{abstract}	

Where particle acceleration and plasma heating take place in relation
to magnetic reconnection is a fundamental question    
for solar flares. We report analysis of an M7.7 flare on 2012 July 19 observed by
\sdoA/AIA and \hsiA. 
{Bi-directional outflows in forms of plasmoid ejections and contracting cusp-shaped loops}
originate between an erupting flux rope and underlying flare loops at speeds of typically
200--$300 \kmps$ up to $1050 \kmps$. 
These {outflows} are associated with spatially separated double coronal X-ray sources 
with centroid separation decreasing with energy.
The highest temperature is located near the nonthermal X-ray loop-top source well below the original 
heights of contracting cusps near the inferred reconnection site.
These observations suggest that the primary loci of particle acceleration and plasma heating 
are in the reconnection outflow regions, rather than the reconnection site itself. 
In addition, there is an initial ascent of the X-ray and EUV loop-top source prior to
its recently recognized descent,	
{ which we ascribe to the interplay among multiple processes
including the upward development of reconnection and the downward contractions of reconnected loops.}
The impulsive phase onset is delayed by 10~minutes from the start of the
descent, but coincides with the rapid speed increases
of the upward plasmoids, the individual loop shrinkages, and the overall loop-top descent,
suggestive of an intimate relation of the energy release rate and reconnection outflow speed.


\end{abstract}

\keywords{acceleration of particles---Sun: flares---Sun: UV radiation ---Sun: X-rays, gamma rays}


\section{Introduction}
\label{sect_intro}



Magnetic reconnection is believed to be the primary energy release mechanism during solar flares,
but where and how the released energy is transformed to heat the plasma and accelerate particles remain 
unclear \cite[for reviews, see][]{HolmanG.HSI.review.2011SSRv..159..107H, FletcherL.HSI.review.2011SSRv..159...19F,
PetrosianV.SA.review.2012SSRv..173..535P, RaymondJ.particle.acceleration.2012SSRv..173..197R}.
Evidence of magnetic reconnection and current sheets on the Sun has been observed in 
various situations and wavelengths. 

A major advance in the last decade was the discovery of a second coronal source
above a commonly observed loop-top source in X-rays and radio wavelengths
\citep{SuiL2003ApJ...596L.251S, SuiL2004ApJ...612..546S, PickM2005ApJ, 
VeronigA2006A&A...446..675V, LiYPGanWQ2007AdSpR,
LiuW_2LT.2008ApJ...676..704L, LiuW.filmnt.2009ApJ...698..632L, 
ChenQR.PetrosianV.2003-11-03-X3.9.2012ApJ...748...33C, SuYang.2nd.heating.2012ApJ...746L...5S, 
BainH.typeII.plasmoid.AIA.2010-11-03.2012ApJ...750...44B, GlesenerL.Krucker.HXR-jet-e-acc.2012ApJ...754....9G}. 
Such double sources often exhibit higher-energy emission being closer to each other, 
indicating higher temperatures or harder spectra of accelerated electrons in the inner region
nearer to the presumable magnetic reconnection site.


Another surprise has been the descent of the loop-top X-ray source at typically 10--$40 \kmps$
early in the impulsive phase before its common ascent through the decay phase
\citep{SuiL2003ApJ...596L.251S, SuiL2004ApJ...612..546S, 	
LiuW2004ApJ...611L..53L, ShenJ.abnormalT.LT.2008ApJ...686L..37S}.
The upper coronal source, however, usually keeps ascending all the time. 

Shrinkages of entire flare loops \citep{SvestkaZ.shrinkage.1987SoPh..108..237S}
at speeds on the order of $10 \kmps$ have been observed in 
soft X-rays \citep[SXRs;][]{ForbesT1996ApJ...459..330F, Reeves.HinodeXRT-shrink.2008ApJ...675..868R},
extreme ultraviolet \citep[EUV;][]{Li&Gan.EUV-shrink.2006ApJ...644L..97L}, and microwaves 
\citep{Li&Gan.radio-shrink.2005ApJ...629L.137L, Reznikova.radio.loop.shrink.2010ApJ...724..171R}. 
They were interpreted as contractions of newly reconnected loops due to magnetic tension
as they evolve from initially cusp shapes toward more relaxed round shapes. 

Often after the impulsive phase, bright loops and dark voids seen in SXR or EUV 
descend onto a flare arcade from above at greater speeds of typically 150$\kmps$ 		
\citep{McKenzieD.HudsonH.SAD-discovery.1999ApJ...519L..93M, 	
SavageS.McKenzieD.SA.downflow.stat.2011ApJ...730...98S}.
At even greater heights of a few $R_\sun$, similar descending loops are seen in white-light coronagraphs,
usually hours after a coronal mass ejection
\citep[CME;][]{WangYM.WL-downflow.1999GeoRL..26.1203W, 	
SheeleyN.WL.TRACE.SADs.2004ApJ...616.1224S}.
These features are also interpreted as contracting post-reconnection loops,
but their observed speeds are only a small fraction of the expected coronal 
\Alfven speed on the order of $\sim$$1000 \kmps$.		

Imaging and Doppler observations have also revealed
bi-directional magnetic reconnection inflows 
\citep{YokoyamaT.reconn.inflow.2001ApJ...546L..69Y, 
MilliganR.plasmoid-LT.inflow.2010ApJ...713.1292M, LiuR.recon.CS.2010ApJ...723L..28L}, 
outflows \citep{InnesD.bi-direction-reconn-outflow.1997Natur.386..811I, KoYK.postCME-CS.2003ApJ...594.1068K,
WangT.SuiL2007ApJ, Nishizuka.SXT.multi.plasmoid.2010ApJ...711.1062N, Hara.reconn.jet.EIS.2011ApJ...741..107H, 
WatanabeT.bi-direction-reconn-outflow-EIS.2012SoPh..tmp..185W},
or both \citep{LinJ.reconn.CS.2005ApJ...622.1251L,
TakasaoS.Asai.recon.in-outflow.2012ApJ...745L...6T, SavageS.recon-in-outflow.2012ApJ...754...13S}.

The physics behind descending X-ray loop-top sources and their relationship with 
slow loop shrinkages, fast supra-arcade loop contractions, and reconnection outflows remain unclear, 
although some models have been proposed \citep[e.g.,][]{Somov.Kosugi.collps.trap.1997ApJ...485..859S,
Karlicky.Kosugi.collps-trap.2004A&A...419.1159K}.	
To fill this gap, we present here observations of a recent eruptive M7.7 flare 		
from the {\it Reuven Ramaty High Energy Solar Spectroscopic Imager}
\citep[\hsiA;][]{LinR2002SoPh..210....3L} and {\it Solar Dynamics Observatory} Atmospheric Imaging Assembly
\citep[\sdoA/AIA;][]{LemenJ.AIA.instrum.2012SoPh..275...17L}.	
We find in this flare all the above interrelated phenomena, which
can be understood in a unified picture as contractions of post-reconnection loops 
modulated by the interplay between energy release and cooling.
The observed highest speed of $\sim$$1000 \kmps$ of loop contractions 
is comparable to the expected \Alfven speed.
The maximum temperature and nonthermal loop-top emission being
away from the inferred reconnection site suggest that primary
heating and particle acceleration take place in the outflow regions,
rather than the reconnection site itself.


After an observational overview in \sect{sect_overview}, we present motions
of the overall X-ray and EUV emission in \sect{sect_motion}.
We examine {bi-directional outflows} in forms of plasmoids and contracting loops
in \sect{subsect_2ejecta}. In \sect{sect_Tmap}, we analyze the spatial distribution of 
energy and temperature dependent emission, including double coronal X-ray sources.	
We conclude in \sect{sect_conclude}, followed by two appendixes 
on supplementary AIA and \stereo observations.
\begin{table}[bthp]	
\scriptsize		
\caption{Event Time Line (2012 July 18--19)}		
\tabcolsep 0.02in	
\begin{tabular}{ll}
\tableline \tableline
 22:18, 07/18	&	Peak of the earlier C4.5 flare	\\
 04:17, 07/19	&	Onset of the M7.7 flare	and initial ascent   \\
                &   of the overall X-ray and EUV loop-top source  \\
\tableline 
 05:02--05:07   &   Onset of overall X-ray and EUV loop-top descent \\
                &   and transition from slow to fast rise of the flux rope CME	  \\ 	
 05:15          &   Max.~velocity ($1050 \kmps$) of upward ejections \\
 05:16          &   Max.~velocity ($-58 \kmps$) of downward loop shrinkages \\
 05:15--05:20   &   Max.~velocity ($-7$ to $-23 \kmps$) of overall  \\
                &   X-ray and EUV loop-top descent \\
 05:16--05:43	&   Flare impulsive phase, hard X-ray burst			\\
 05:21--05:31   &   Min.~height of overall X-ray and EUV loop-top descent \\ 
                &   and onset of the second ascent   \\
\tableline 
 04:17--05:20   &   Upward ejections; downward, low-altitude fast contractions \\
 05:20--16:00   &   Downward, high-altitude fast loop contractions  \\
 04:17--16:00   &   Downward slow loop shrinkages  \\
 06:52          &   Max.~velocity ($-918 \kmps$) of downward fast contractions  \\ 

\tableline  \end{tabular}

\label{table_timeline} \end{table}
%

\section{Overview of Observations}
\label{sect_overview}

The event under study was an M7.7 flare that occurred at $\sim$04:17~UT on 2012 July 19
in NOAA active region (AR) 11520 on the southwest limb. It was well observed by \hsi and \sdoA/AIA,
but it was not detected by \fermi and its impulsive phase was missed by the X-ray Telescope on \hinodeA.
\tab{table_timeline} summarizes the event time line that will be discussed in detail.
 \begin{figure}[thbp]      
 \epsscale{1.1}	
 \plotone{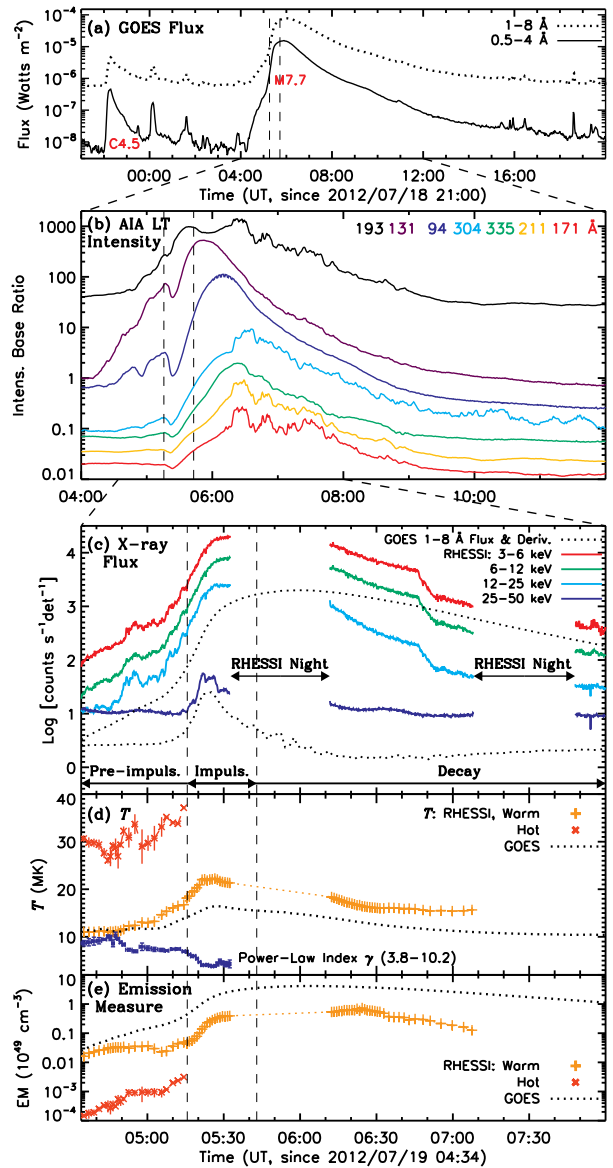}
 \caption[]{\footnotesize
 History of X-ray and EUV flare emission.
  (a) \goes SXR fluxes showing the preceding C4.5 flare and the main M7.7 flare.
  (b) \sdoA/AIA intensity of the loop-top region 
 (at projected height $h_{\rm ref}=60\arcsec$ on Cut~0, as shown in  \fig{tslice_all.eps}) 
 normalized by its initial value (base ratio). Color-coded for AIA channels, the curves
 are vertically shifted to avoid overlap and arranged from the top to bottom approximately 
 in the temporal order of their initial response to the flare 
 (in progressively cooler channels).
  (c) \hsi count rates in colored solid lines
 and \goes 1--8~\AA\ flux and its time derivative in black dotted lines,
 arbitrarily shifted vertically.	
 The two vertical dashed lines indicate the impulsive phase.
  (d) Temperature and (e) emission measure of the flare plasma inferred from  
 \hsi (orange/red) and \goes (black) spectral fits, 
 together with the power-law index $\gamma$ of the nonthermal component shown in blue in (d). 
 } \label{lc.eps}
 \end{figure}
%

\fig{lc.eps} shows the history of the flare emission. The \goes 1--8~\AA\ flux peaks at 05:58~UT followed by 
a slow decay lasting almost one day.
We define the interval of 05:16--05:43~UT 	
as the impulsive phase, as marked by the two vertical dashed lines, 
which starts at the sudden rise of the \hsi 25--50~keV flux and ends (during \hsi night)
when the time derivative of the \goes 1--8~\AA\ flux drops to its level at the impulsive phase onset,
assuming the \citet{NeupertW1968ApJ...153L..59N}	
effect at work. We call the intervals before and after the impulsive phase 
the pre-impulsive and decay phases. \hsi has good coverage except for the late
impulsive and early decay phases.

\begin{figure*}[thbp]      
 \epsscale{0.95}	
 \plotone{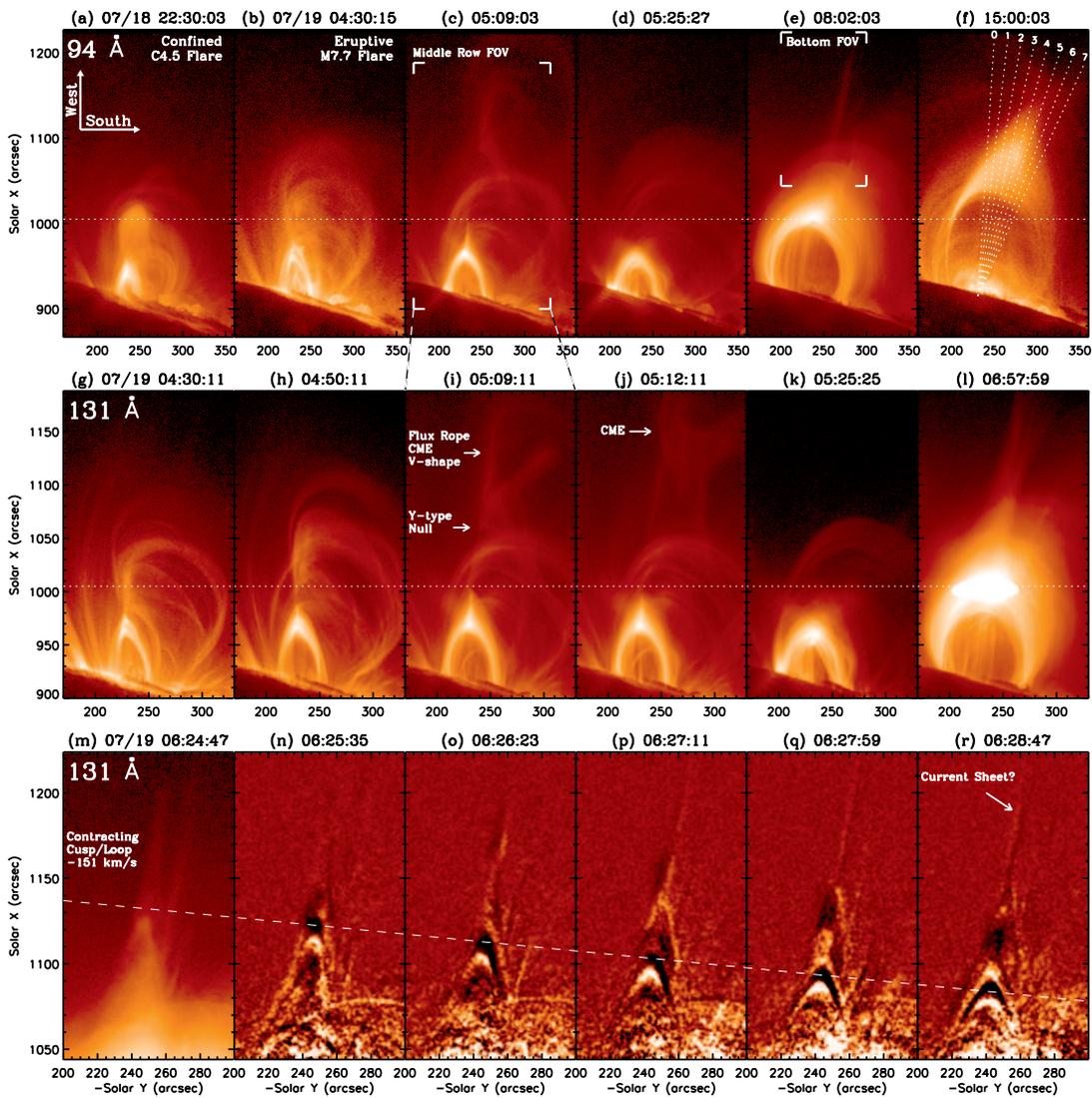}
 \caption[]{\footnotesize
 AIA images of the event, rotated to the solar west up.
   Top: 94~\AA\ image sequence of the earlier C4.5 flare and the M7.7 flare under study
 (see the associated online Movie~A for composite 94 (red) and 335~\AA\ (green) images).
 The brackets in (c) and (e) mark the enlarged fields of view (FOV) of the
 middle and bottom panels, respectively. The numbered dotted lines in (f) mark the cuts 
 for obtaining space-time plots presented in this paper.
   Middle: Detailed 131~\AA\ image sequence showing the upward growth of the flare cusp,
 followed by its rapid shrinkage around the	
 early impulsive phase. 
 The  horizontal dotted line marks the highest position of the early cusp.
   Bottom: 131~\AA\ image sequence showing an example of pointed cusps contracting
 from a ray-like structure, presumably a reconnecting current sheet
 (see the associated Movie~B).	
 The first panel is an original image and the rest are running difference images.
 The slanted dashed line indicates an average contraction velocity of $-151 \kmps$.
 } \label{maps_overview.eps}
 \end{figure*}

\fig{maps_overview.eps} and its associated Movie~A show AIA images of the event. An earlier C4.5 flare occurred 
in the same location, peaking at 22:18~UT on 2012 July 18 (see \fig{lc.eps}(a)).
This is a {\it confined} flare that produces a hot flux rope failing to erupt and cusp-shaped flare
loops underneath it (\fig{maps_overview.eps}(a)). 
This configuration then gradually evolves for hours and finally becomes unstable,
initiating the later, {\it eruptive} M7.7 flare, when
the flux rope is expelled as a fast CME of $>$$1000 \kmps$.
The flux rope in this two-stage eruption was reported by 
\citet{Patsourakos.pre-exist-flux-rope_CME.2013ApJ...764..125P}.

As shown in \fig{maps_overview.eps}(i), the trailing edge of the CME displays a clear ``V-shape",
which, together with the underlying ``inverted-V shape" of cusp-like flare loops, 
suggests two Y-type null points with a vertical current sheet formed in between, 
as predicted in the classical picture of eruptive flares 
\citep{CarmichaelH1964psf..conf..451C, SturrockP1966Natur.211..695S, 
HirayamaT1974SoPh...34..323H, KoppR1976SoPh...50...85K}.
Not predicted in that picture is	
the initial upward growth of the cusp
followed by its rapid downward shrinkage around the early impulsive phase,
prior to another, expected upward growth through the decay phase in a 
commonly observed candle-flame shape (\figs{maps_overview.eps}(a)--(l)).	
The initial growth and shrinkage are accompanied by the gradual rise and
impulsive eruption of the overlying flux rope, respectively. 
Equally interesting are high-speed {bi-directional outflows} involving
upward-moving plasmoids	
and downward-contracting pointed cusps  (\fig{maps_overview.eps}, bottom; \fig{maps_blob.eps}).   
We examine these and related features observed by \hsi and AIA in next several sections.
 \begin{figure*}[thbp]      
 \epsscale{1.0}	
 \plotone{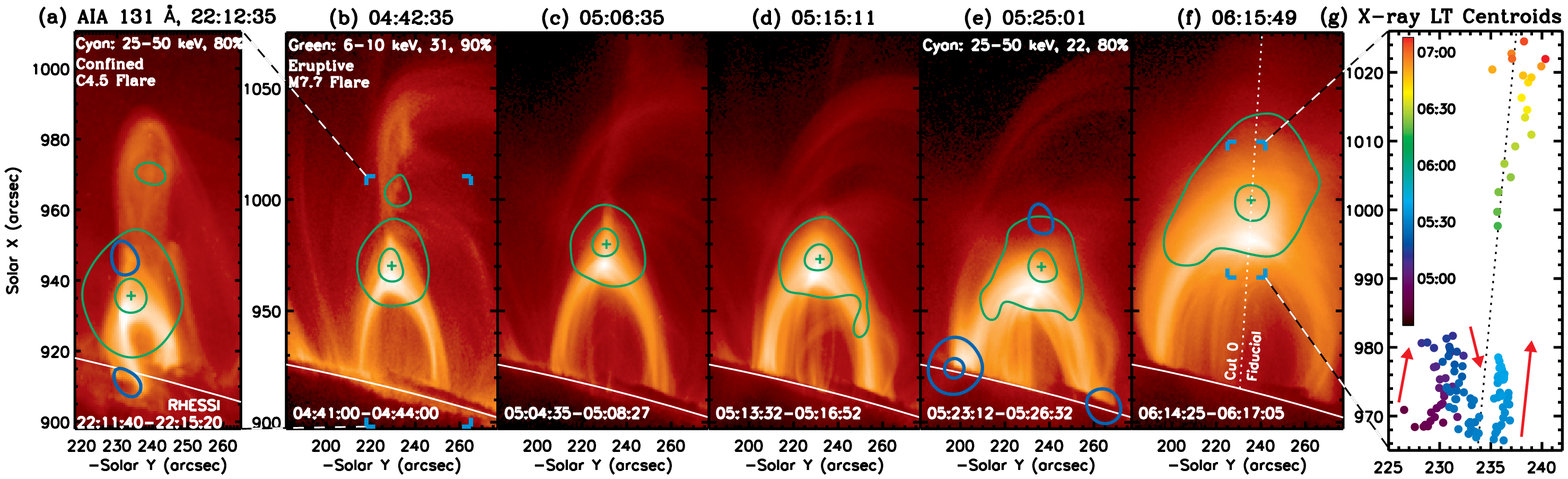}
 \caption[]{\footnotesize
  (a)--(f) Evolution of \hsi 6--10~keV sources (green contours at 31\% and 90\% of the maximum)
 overlaid on concurrent AIA 131~\AA\ images (see the accompanying Movie). 
 The blue contours show 25--50~keV emission, for both the C4.5 and M7.7 flares, 
 at the footpoint(s) and the Masuda-type coronal source $20\arcsec$ above the SXR and EUV loops.
  (g) Color-coded temporal migration of the 6--10~keV loop-top centroid as marked by the plus sign
 on the left. The red arrows indicate an initial ascent followed by a descent 
 and then another ascent. 	
 The dotted line marks the fiducial line (Cut~0) for obtaining projected heights shown in \fig{cent_time.eps}.
 The first and last panels are enlarged from the bracketed regions shown in
 their adjacent panels.
 } \label{maps_hsi.eps}
 \end{figure*}
%

\section{Overall X-ray and EUV Source Motions}	
\label{sect_motion}

We first follow the evolution of the morphologies and positions of overall
X-ray and EUV sources.

\subsection{X-ray Source Morphology}
\label{subsect_XrayMotion}

We reconstructed \hsi X-ray images in energy bands from 3 to 50~keV
using the CLEAN algorithm and detectors 3--9.
Depending on the count rate, we chose variable integration time ranging from 20~s during the impulsive phase to
4~minutes during the decay phase.	

\fig{maps_hsi.eps} and its accompanying Movie show examples of \hsi contours overlaid on AIA images. There is a persistent
loop-top source at 6--10~keV (green) near the apex of the cusp-shaped EUV loops. Accompanying the 
evolution of the EUV loops mentioned above, this loop-top source undergoes a gradual ascent
followed by a descent and then another ascent. This can be best seen in the last panel
showing the temporal migration of the emission centroid obtained from a contour at 70\% of the maximum of each image. 
Early descents of loop-top X-ray sources have been recognized \citep[e.g.,][]{SuiL2004ApJ...612..546S}, 
but this is the first time that an evident preceding ascent is observed. 

Early in the event there is an additional, weaker 	
coronal X-ray source (panel (b)) located in the lower portion of the overlying flux rope. 
It later falls below detection with the eruption of the flux rope.
During the impulsive phase (panel (e)) at higher energies (25--50~keV), there are double footpoint sources 
and a Masuda-type \citep[e.g.,]{Masuda1994Nature, NittaN.revisit-Masuda.2010ApJ...725L..28N} 
coronal source located $20\arcsec = 15 \Mm$ above the
SXR (6--10~keV) loop-top. This separation is twice that of the Masuda case.

The earlier C4.5 flare (panel (a)), though more compact, exhibits surprisingly similar emission,
including the additional X-ray source within the flux rope and the Masuda-type hard X-ray (HXR) source.
This suggests that the end state of the first, confined flare of a failed eruption serves as 
the initial state of the second, eruptive flare, making them homologous flares.
Note that the weaker, southern footpoint emission is (partially) occulted by the limb,
especially for the first flare.

 \begin{figure}[thbp]      
 \epsscale{1.2}	
 \plotone{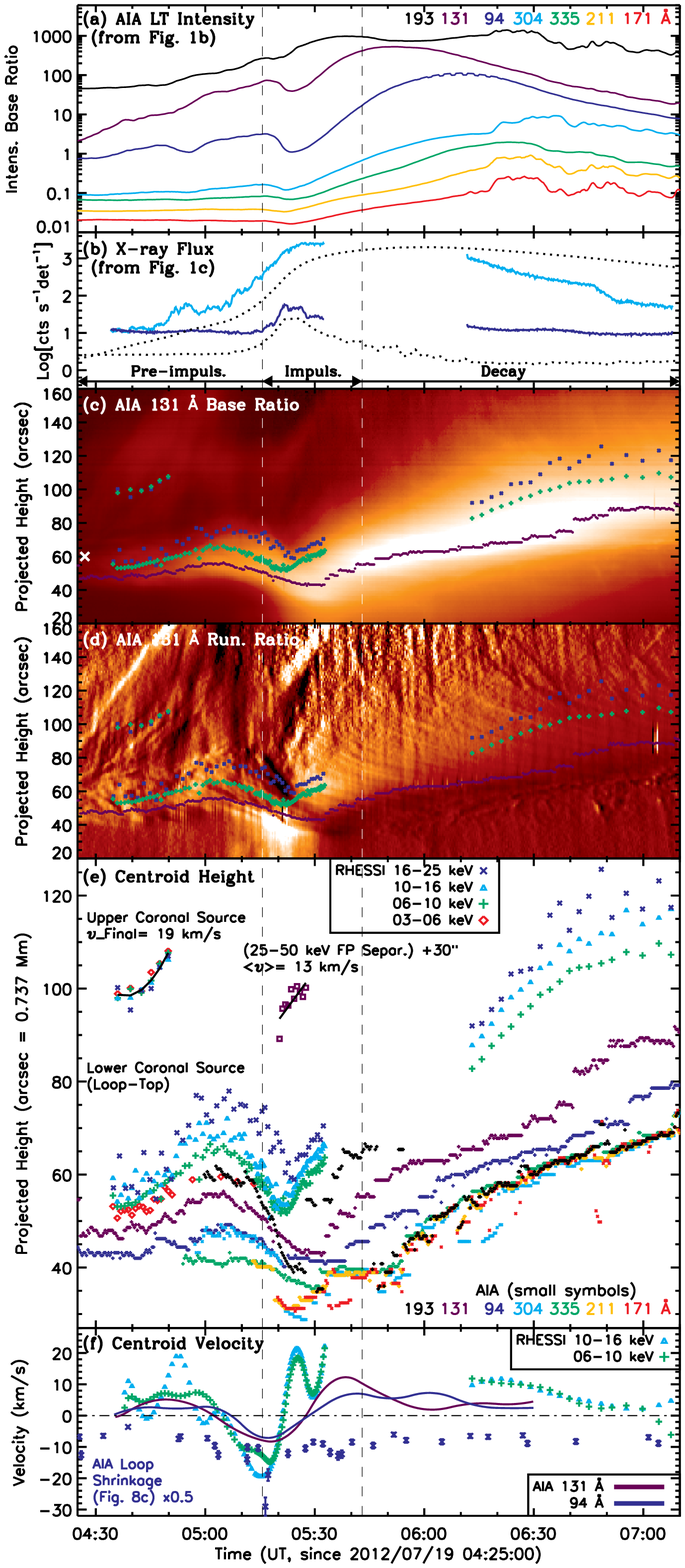}
 \caption[]{\footnotesize
 X-ray and EUV source motions.	
  (a) and (b) Subsets of light curves from \figs{lc.eps}(b) and (c).
  (c) and (d) Base and running ratio space-time plots of the AIA 131~\AA\ channel from Cut~0,
 overlaid with selected centroid positions from (e).
  (e) Projected heights of the emission centroids of the upper and lower coronal sources from \hsi (large symbols, greater heights)
 and of the loop-top peaks from AIA (small symbols, lower heights).
 Also shown is the 25--50~keV footpoint separation (purple squares) shifted upward by $30\arcsec$.
  (f) Velocity of centroids from spline fits to selected channels in (e). The blue symbols are
 the initial velocities of individual 	
 but rescaled by a factor of 0.5.
 } \label{cent_time.eps}
 \end{figure}
%

\subsection{X-ray and EUV Height-time History}      
\label{subsect_ht}

To track various moving features, we placed eight cuts of $10\arcsec$ wide centered on the limb, 
as shown in \fig{maps_overview.eps}(f). Cut~0 is positioned along the 
presumable current sheet between the initial flare cusp and erupting flux rope,
and is used as the fiducial line for obtaining projected heights up to $\sim$08:00~UT.
Subsequent cuts are evenly spaced by $3\degree$ to capture the cusp at different stages 
as it gradual turns toward the south. For each cut, we averaged pixels of an AIA image sequence 
within it in the perpendicular direction to obtain a space-time plot.

\fig{cent_time.eps}(c) shows, for example, a base-ratio (i.e., normalized by the initial intensity profile) 
space-time plot from Cut~0 at 131~\AA. It shows the eruption of 
the flux rope and the three-stage development (upward, downward, and upward again) of the underlying 
flare cusp. The peak emission at each time, marked by the small purple symbols,
evidently exhibits this motion. Space-time plots of other AIA channels are
shown in \fig{tslice_all.eps} and described in Appendix~A.

\fig{cent_time.eps}(e) shows the height-time history of the loop-top emission centroids
in four \hsi X-ray bands from 3 to 25~keV and in all AIA EUV channels.
(From now on, we refer to both X-ray centroids and EUV peaks as centroids.)
We find that all loop-top centroids follow the same three-stage motions.
The AIA 131 and 94~\AA\ channels best exhibit this trend continuously,
while in cooler channels, especially 211 and 171~\AA, 
the initial ascent and subsequent descent become obscure.
Note that the 3--6~keV centroid (red diamonds) cannot be identified since the impulsive phase because
the \hsi thin attenuator moves in and raises the energy threshold to 6 keV.
 \begin{table*}[thbp]      
 \scriptsize
 \caption{Projected Heights and Velocities of \hsi and AIA Loop-top (LT) Centroids}
 \tabcolsep 0.05in
 \begin{tabular}{rrrrcrrrrrrrrrr}
 \tableline\tableline
 Channels & \multicolumn{4}{c}{Height (arcsec=0.737 Mm)/Time (UT)}  && \multicolumn{7}{c}{Velocity ($\kmps$)/Time (UT)}   \\
            \cline{2-5}                                          \cline{7-13}
           & Initial               & Max  & Min  & Descent \% && Initial Ascent & \multicolumn{2}{c}{Descent}   && \multicolumn{3}{c}{2nd Ascent}   \\
                                                                                 \cline{8-9}                     \cline{11-13}
                                       & ($\sim$04:35) & & &  &&        & Mean &  Max                   &&  --6:00  & 6:00--7:00  & 7:00--8:00  \\
 \tableline
 \multicolumn{1}{c}{\hsi} & & &  &&  & & &&  &  & \\

 16--25 keV & 59\arcsec & 77\arcsec/05:07 & 60\arcsec/05:24 & 22\%  &&  8.9 & $-12$ & $-23$/05:20 && 14  & 12  & 1.5  \\
 10--16 keV & 55\arcsec & 69\arcsec/05:06 & 54\arcsec/05:21 & 22\%  &&  7.1 & $-15$ & $-19$/05:15 && 14  & 10  & 3.6  \\
 6--10 keV  & 53\arcsec & 66\arcsec/05:05 & 52\arcsec/05:22 & 21\%  &&  6.0 & $-11$ & $-15$/05:18 && 13  & 9.6 & 2.1  \\
 3--6 keV   & 52\arcsec & 60\arcsec/05:04 &     ...  & ... &&  4.1 & ... & ...       && ... & ... & ...  \\

 \tableline
 \multicolumn{1}{c}{AIA}  & &  &  &&  & & && & & \\

 193 \AA & ...        & 61\arcsec/05:03 & 35\arcsec/05:31 & (43\%) && ... & $-13$  & $-20$/05:19  && ... & 4.9 & 3.0 \\
 131 \AA & 47\arcsec  & 56\arcsec/05:02 & 43\arcsec/05:30 & 23\%   && 4.3 & $-5.9$ & $-8.3$/05:18 && 11  & 3.6 & 2.6 \\
 94  \AA & 43\arcsec  & 49\arcsec/05:06 & 41\arcsec/05:29 & 16\%   && 2.5 & $-4.7$ & $-7.1$/05:17 && 5.9 & 3.2 & 2.6 \\

 304 \AA & ...        & 48\arcsec/05:04 & 29\arcsec/05:27 & (40\%) && ... & $-5.1$ & ...        && 4.6 & 4.1 & 2.4 \\
 335 \AA & ...        & 41\arcsec/05:07 & 36\arcsec/05:29 & 12\%   && ... & $-3.9$ & ...        && 4.1 & 3.5 & 3.2 \\

 211 \AA & ...        & ...      & 31\arcsec/05:23 & ...  && ... & $-6.9$ & ...        && 5.0 & 4.3 & 3.2 \\
 171 \AA & ...        & ...      & 31\arcsec/05:24 & ...  && ... & $-6.0$ & ...        && 5.6 & 5.0 & 3.5 \\

 \tableline
 \end{tabular}
 \tablecomments{Centroid height uncertainties are $\lesssim$$2\arcsec$ for \hsi and $\lesssim$$1\arcsec$ for AIA.
 Velocity uncertainties are on the order of 10\%--15\%.}
 \label{table_cent}
 \end{table*}
%
%
%

We also notice a clear energy dispersion that the loop-top emissions at higher temperatures 
or photon energies are located at greater heights,
with all X-ray centroids situated above EUV centroids.
For example, at its greatest height prior to the impulsive phase, the 16--25~keV centroid
is $35\arcsec= 26 \Mm$ above its 335~\AA\ counterpart.
This energy dispersion is consistent with previous observations 
\citep[e.g.,][]{VeronigA2006A&A...446..675V}	
and in line with the expected picture that higher loops are newly energized 
and thus are hotter and/or host nonthermal electrons of harder spectra, 
while lower loops are previously energized and have undergone cooling. 
We suggest that the so-called above-the-loop-top Masuda-type sources, 
including the 25--50~keV coronal source in \fig{maps_hsi.eps}(d), could be extreme cases of this general trend.

In \tab{table_cent}, we list the initial heights of X-ray and EUV loop-top centroids around 04:35~UT,
together with the maximum and minimum heights during their descents and the percentage height reductions.
In general, the above noted energy dispersion persists and the 
10--20\% descents are in agreement with earlier reports \citep{VeronigA2006A&A...446..675V}.
An exception is the 193 and 304~\AA\ channels because of their response to both hot and cool emissions
(see Appendix~A).	
There is a weak trend that higher energy X-ray descents start (at the maximum heights)
later, but still within 5~minutes during 05:02--05:07~UT,
and X-ray descents end (at the minimum heights) earlier than EUV descents
by up to 10~minutes.

We measured the average velocities of the loop-top centroids by fitting
a piecewise linear function to the height-time data during their initial ascent,
subsequent descent, and second ascent stages.
As listed on the right hand side of \tab{table_cent},
the result shows a general trend of higher velocities at higher energies.
The X-ray descents of $-11$ to $-15 \kmps$ are about twice faster than 
the EUV descents except for 193~\AA\ because of its dual temperature response. 
There is also a gradual decrease in velocity with time
during the second ascent stage through the decay phase.
These trends generally agree with previous observations 
\citep{LiuW2004ApJ...611L..53L, VeronigA2006A&A...446..675V}.

To follow the temporal variations of the loop-top velocities, we took time derivatives of spline fits
to the centroid heights of selected channels.
The resulting velocities vs.~time are shown in \fig{cent_time.eps}(c)
for the 6--10 \& 10--16~keV and 131 \& 94~\AA\ channels.
The outstanding feature is that the maximum descent velocities
(also tabulated in \tab{table_cent}) in the range of $-7$ to $-23 \kmps$ 
are all attained during 05:15--05:20~UT, 
near the onset of the impulsive phase at 05:16~UT marked by a vertical dashed line.
This time is delayed from the initial descent by about 10~minutes. 
This indicates that the loop-top descent velocity, rather than
its height, is more intimately correlated with the energy release rate.

%
 \begin{figure}[thbp]      
 \epsscale{1.1}	
 \plotone{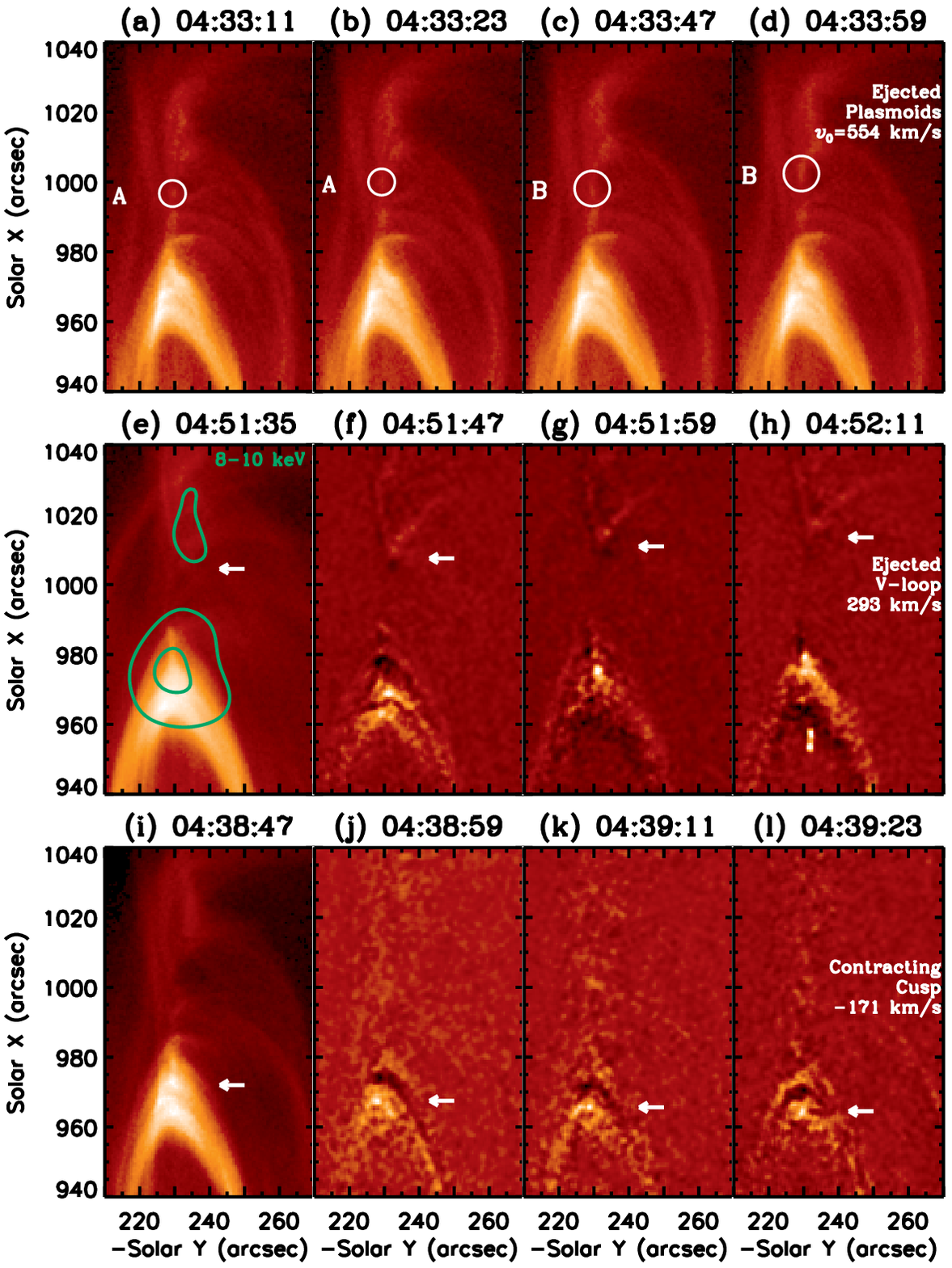}
 \caption[]{\footnotesize
 AIA 131~\AA\ images (see the accompanying Movie) 	
 showing upward moving blobs (top, labeled A and B)		
 and V-shapes (middle) and downward contracting cusps (bottom).
 Their space-time fits are shown in \fig{tslice_SAD.eps}(a).
 The green contours show two distinct 8-10~keV \hsi sources from \fig{cent_E.eps}(c).
 The first row and column are original images and the rest are running difference images.
 } \label{maps_blob.eps}
 \end{figure}
%
We also show in \fig{cent_time.eps}(e) the centroid heights of the upper coronal X-ray source
located in the lower portion of the flux rope.
A parabolic fit 	
indicates a final velocity of $19 \kmps$ at 04:50~UT, comparable to or slower than those in
other events \citep{LiuW_2LT.2008ApJ...676..704L, LiuW.filmnt.2009ApJ...698..632L, SuiL2003ApJ...596L.251S}.
It is also 50\% slower than the middle portion of the then accelerating flux rope, 
as indicated by the parabolic fit in \fig{tslice_all.eps}(e), 
which reaches $153 \pm 3 \kmps$ later at 05:07~UT near the edge of AIA's FOV.

The conjugate footpoints at 25--50~keV during the impulsive phase move away from each other
at an average velocity of $13 \kmps$ (see \fig{cent_time.eps}(e)), almost identical to
the velocity of the simultaneous loop-top ascent (see \tab{table_cent}).
This is consistent with previous observations \citep{LiuW2004ApJ...611L..53L}	
and the classical picture of eruptive flares during the arcade growth phase.

\section{{Bi-directional Plasma Outflows}}		
\label{subsect_2ejecta}	

We now turn our attention to the {bi-directional plasma outflows} observed by AIA.   
As shown in \fig{maps_blob.eps} and its accompanying Movie, emission features {move}	
both upward and downward from above the cusp-shaped flare loops.
The upward {outflows}, observed only when the flux rope rises toward its eruption,
are mainly bright blobs (interpreted as plasmoids, top row),	
while the downward {outflows} are primarily in the form of retracting cusp-shaped loops (bottom row)
observed throughout the event well into the decay phase. Among the downward retracting
loops, those seen at low altitudes tend to have lower speeds ($\lesssim$$60 \kmps$)
and gradually decelerate when approaching and piling up onto the apex of the flare arcade, 
while those originating from high altitudes tend to have higher speeds ($\gtrsim$$100 \kmps$)
and decelerate more rapidly. According to these observational distinctions, we call the former
{\it slow loop shrinkages} and the latter {\it fast loop contractions}, although they might share
a common physical origin as discussed in \sect{sect_conclude}. 
The former are likely analogous to shrinkages observed at SXR and other wavelengths
\citep{SvestkaZ.shrinkage.1987SoPh..108..237S}, while the latter, especially those occurring after the flare arcade has formed, 
are likely so-called supra-arcade downflow loops \citep	
{SavageS.McKenzieD.SA.downflow.stat.2011ApJ...730...98S}.
 \begin{figure*}[thbp]      
 \epsscale{1.0}	
 \plotone{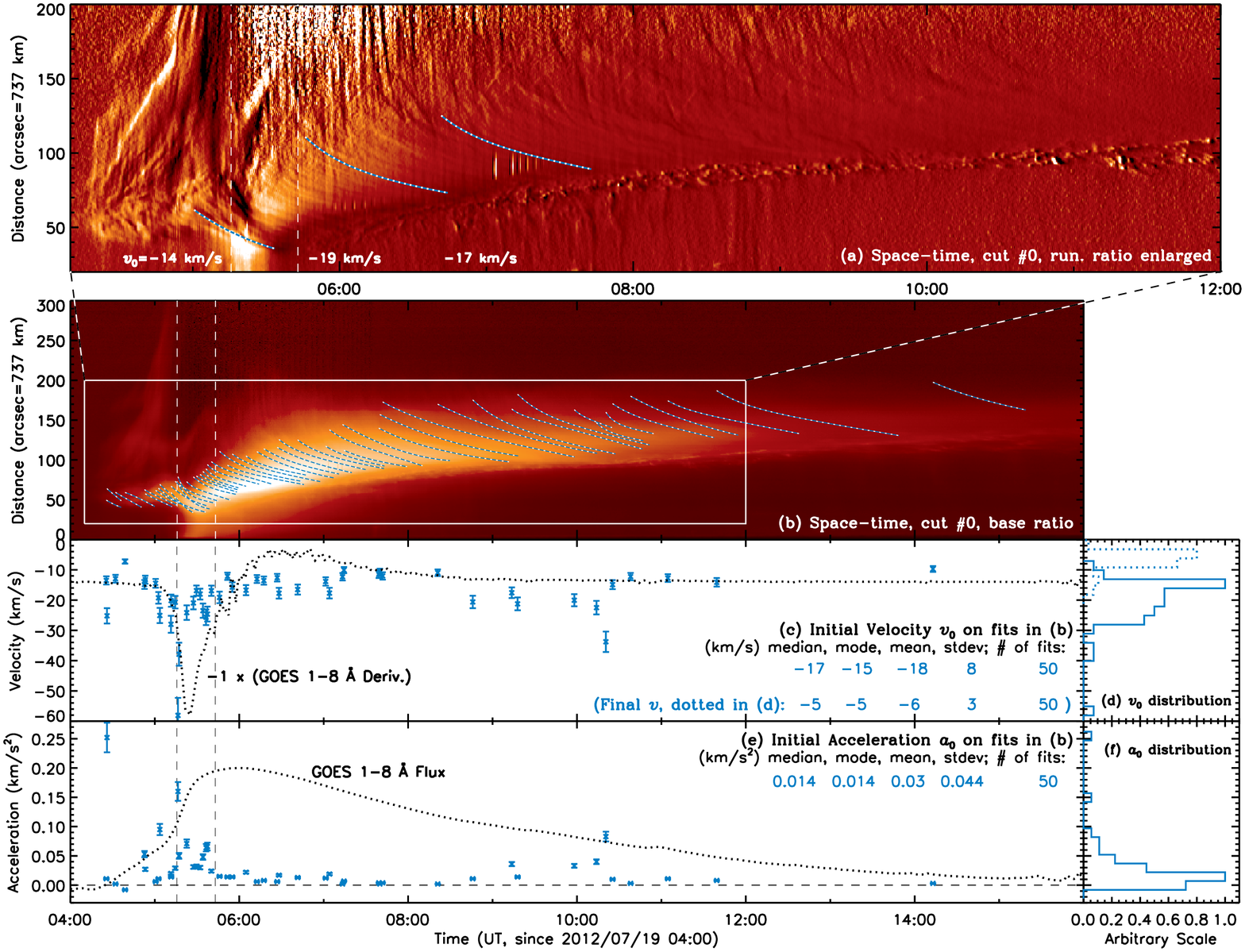}
 \caption[]{\footnotesize
 Kinematic measurements of slow loop shrinkages.	
  (a) Running ratio space-time plot enlarged for the boxed region in (b),
 overlaid with examples of fits to shrinkage tracks labeled with their initial velocities.
  (b) Base ratio 131~\AA\ space-time plot from Cut~0 overlaid with all 50 fits performed.
  (c) Initial velocities $v_0$ of the fitted tracks in (b) as a function of time.
  (d) Histograms of $v_0$ (solid) and the final velocity $v_f$ (dotted)
 measured at the end of each track, with their statistical properties
 (median, mode, mean, standard deviation, and total number of fits) listed in (c).
  (e) and (f) same as (c) and (d), but for the initial acceleration $a_0$.
 The black dotted lines in (c) and (e) are the time derivative (inverted)
 and original \goes 1--8~\AA\ flux.
 The two vertical dashed lines mark the impulsive phase.
 } \label{tslice_shrink.eps}
 \end{figure*}
%

To track these moving features, we obtained space-time plots 
from the cuts defined in \fig{maps_overview.eps}(f) using the 131~\AA\
channel because it provides the best coverage of these features. We then applied running ratio,
namely, dividing the intensity profile at each time by its previous neighbor,
which highlights moving features as intensity tracks (see, e.g., \fig{cent_time.eps}(d)).
We exhaustively identified such tracks 
of more than $10\arcsec$ long corresponding to unique moving features.
If one feature is captured by multiple cuts, we included only the most complete track.
Following \citet{WarrenH.AIA.SADs.2011ApJ...742...92W}, we fitted the projected height $h$ 
of each track by a function of time $t$,
 \beq
  h(t)= h_0 + v_T t + a_0 \tau^2 (e^{-t/\tau} - 1),
 \eeq \label{hfit.eq}
which gives the instantaneous velocity
 \beq
  v(t) = v_T - a_0 \tau e^{-t/\tau},
 \eeq \label{vfit.eq}
and acceleration
 \beq
  a(t) = a_0 e^{-t/\tau},
 \eeq \label{afit.eq}
where $h_0$ and $a_0$ are the initial height and acceleration,
$v_T$ is the terminal velocity, and $\tau$ is the $e$-folding decay time	
that characterizes the observed decreasing acceleration 
(or deceleration) with time. We present below the kinematics of the features
categorized above and focus on their initial velocities $v_0= v_T - a_0 \tau $ and accelerations $a_0$.


\subsection{Slow Downward Loop Shrinkages}
\label{subsect_shrink}

 \begin{figure*}[thbp]      
 \epsscale{1.0}	
 \plotone{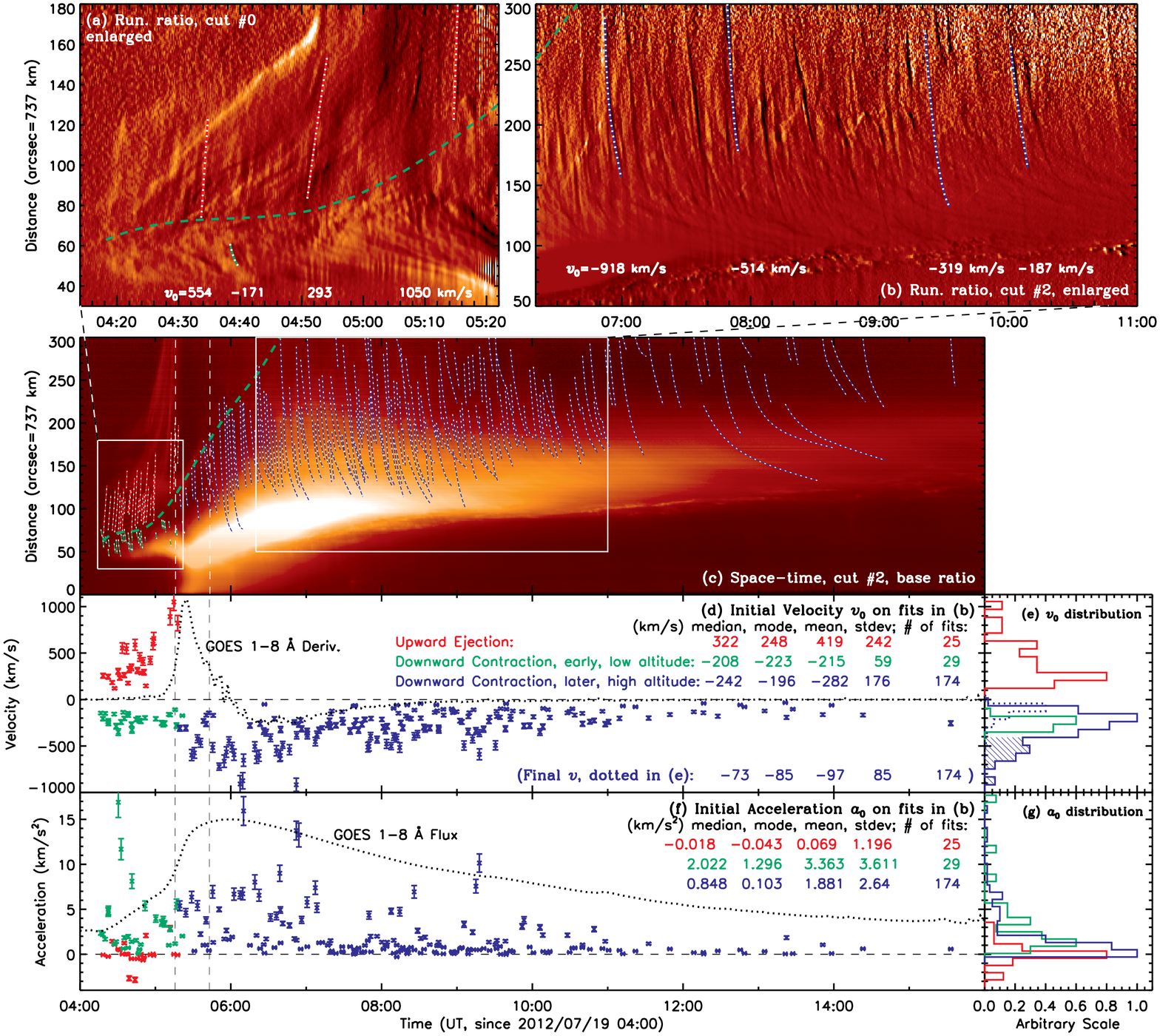}
 \caption[]{\footnotesize
 Same as \fig{tslice_shrink.eps} but for fast { upward plasmoid ejections and downward loop contractions}.	
   (a) and (b) Running ratio space-time plots obtained from Cuts~0 and 2,
 enlarged for the two boxed regions in (c).
   Overlaid with examples of color-coded fits, (a) shows simultaneous upward ejections (red dotted)
 and downward contractions of early, low-lying loops (green dotted), and (b) shows downward contractions 
 of later, high-lying loops (blue dotted) from well above the flare arcade.
 Example images corresponding to the fits in (a) and (b) are shown in \figs{maps_blob.eps}
 and \ref{maps_overview.eps} (bottom), respectively.
  (c) Base ratio 131~\AA\ space-time plot from Cut 2 overlaid with all fits to these identified features.
 The thick green dashed line is a spline fit to the initial heights of { bi-directional outflows},	
 as an estimate to the position of reconnection.
  (d)--(g) Initial velocities $v_0$ and accelerations $a_0$ of the color-coded fits in (c) as a function of time,
 together with their histograms on the right.
 The blue dotted line in (e) is for the final velocities of the late, high-lying loop contractions.
 The short, vertical stripes in (a)  	
 with a periodicity of $\sim$1~minute between 04:45--05:15~UT are possibly due to quasi-periodic flare pulsations 
 \citep{Nakariakov.Melnikov.QPP.2009SSRv..149..119N},
 not to be confused with the tracks of steep slopes of { plasma outflows}.	
 } \label{tslice_SAD.eps}
 \end{figure*}
%
%
\fig{tslice_shrink.eps} shows kinematic measurements of the slow loop shrinkages.
These loops traverse a hot (up to $T \sim 30 \MK$; see \fig{tem_map.eps}) 
loop-top region, which is bright only in the hottest channels of 
193, 131, and 94~\AA, and descend toward the apex of the flare arcade seen
in cooler channels, where they fade below detection (see \fig{tslice_all.eps}). 
They generally start at a low velocity with a median of $-17 \kmps$ ($+/-$ for upward/downward) 
and decelerate to typically $-5 \kmps$ within 0.5--2.5 h. Their median initial deceleration is $0.014 \kmpss$.

Such shrinkages are persistent throughout the flare,
not only during the decay phase when the overall loop-top emission develops upward
as previously observed \citep{ForbesT1996ApJ...459..330F},	
but also during the pre-impulsive and impulsive phases when the loop-top
undergoes its initial ascent and subsequent descent.
50 shrinkages are identified during 04:00--16:00~UT with an average
occurrence rate of once per 7.7~minutes up to 10:00~UT
but more frequent around the impulsive phase (\fig{tslice_shrink.eps}(b)).

Regardless of the direction of motion of the loop-top emission,
the initial height of loop shrinkage generally increases with time.
The initial velocity $v_0$ exhibits more variability, as shown in \fig{tslice_shrink.eps}(c),
and appears to be positively correlated with the flare energy release rate
indicated by the time derivative of the \goes flux (black dotted line).
In particular, near the onset of the impulsive phase at 05:16~UT (vertical dashed line),
$v_0$ increases rapidly and reaches a maximum of $-58 \kmps$.
There is a similar temporal correlation with the initial deceleration $a_0$ (panel (d))
that also generally correlates with $v_0$.

\subsection{Fast Downward Loop Contractions}
\label{subsect_SAD}

\fig{tslice_SAD.eps} shows, in the same form as \fig{tslice_shrink.eps}, 
kinematic measurements of fast downward loop contractions
as well as upward plasmoid ejections that will be examined in \sect{subsect_upward}.
We color-code fits to tracks of different categories: upward ejections
in {\it red} and downward contractions in cooler colors.
For the latter, we use {\it green} for early contractions that occur simultaneously with
upward ejections and originate from lower heights $h<100\arcsec$,
and {\it blue} for later contractions that occur without observed upward counterparts
(possibly out of AIA's FOV, after the flux rope has erupted)
and originate from greater heights $h>100\arcsec$.
We identified 29 green tracks and 174 blue tracks during 04:00--16:00~UT.

As shown in \fig{tslice_SAD.eps}(c), the fast downward contractions, especially those later blue-colored ones,
start well above the hot loop-top region and travel into it with deceleration for some distance
before fading below detection. Their final heights are generally above the original heights 
of slow loop shrinkages (see \fig{tslice_shrink.eps}(b)). 

In original AIA images, the fast contracting features are usually bright, cusp-shaped loops
(see \fig{maps_overview.eps}, bottom),
but occasionally dark, tadpole-like voids 
\citep{McKenzieD.HudsonH.SAD-discovery.1999ApJ...519L..93M, InnesD.SADs.void.2003SoPh..217..247I,
CassaP.DrakeJ.SADs.low-density-recon}.
In running ratio space-time plots (\fig{tslice_SAD.eps}, top), 
the former each produce a bright track followed by a recovering dark track,
while the latter produce tracks of reversed order.
We treat them all as contracting loops and do not distinguish their 
differences that are beyond our scope.
 \begin{table*}[thbp]      
 \tiny	
 \caption{Kinematic Statistics of { Bi-directional Outflows}}
 \tabcolsep 0.02in
 \begin{tabular}{lcrrrrrrrrrrrrrrrrrrrrrrrr}
 \tableline\tableline
 Features   & Interval$^{(a)}$ & \multicolumn{5}{c}{Initial Velocity ($\kmps$)}                    && \multicolumn{5}{c}{Final Velocity ($\kmps$)}  && \multicolumn{5}{c}{Initial Acceleration ($\kmpss$)}  \\
                        \cline{3-7}                                                  \cline{9-13}                                     \cline{15-19}
 (\# of Fits) & (minutes)  & Med.  &  Mod.  & $\sigma$  & Max/Time  & Min           && Med.  &  Mod.    & $\sigma$  & Max/Time  & Min  && Med.  &  Mod.    & $\sigma$  & Max/Time  & Min  \\
 \tableline
 Ejected Plasmoids (25) & 2.4 &$ 322 $&$ 248 $&$  242 $&$ 1050$/{\bf 05:15} &$  121 $&&$  343  $&$ 320 $&$  250  $&$ 1070 $/05:13 &$ 70 $&&$ -0.018 $&$  -0.043 $&$  1.2   $&$ 2.89$/05:12 &$ -2.84 $ \\
 Shrinking Loops (50)   & 7.7 &$ -17 $&$ -15 $&$   8  $&$ -58$/{\bf 05:16}  &$  -7  $&&$  -5   $&$  -5 $&$   3  $&$ -18$/04:26 &$ -2    $&&$  0.014 $&$  0.014  $&$ 0.044  $&$ 0.252$/04:26 &$-0.008 $ \\
 Contracting Cups: \\
 Early, Low Altit.(29)  & 2.2 &$ -208 $&$ -231 $&$  59  $&$ -351$/04:30  &$ -110 $&&$ -89  $&$ -68   $&$ 58  $&$ -226$/04:30 &$ -26  $&&$  2.022  $&$  1.30   $&$  3.61   $&$ 16.9$/04:30 &$ 0.096 $ \\
 Late, High Altit.(174) & 2.0 &$ -242 $&$ -158 $&$ 176  $&$ -918$/06:52 &$ -34  $&&$ -73  $&$ -113   $&$ 85  $&$ -553$/05:32 &$ -1   $&&$  0.848 $&$  0.103  $&$  2.64   $&$ 15.9$/06:10 &$ -0.297 $ \\

 \tableline
 \end{tabular}
 \tablecomments{Abbreviations: Med. for Median, Mod. for Mode. $^{(a)}$ Average occurrence interval up to 10:00~UT.}
 \label{table_vstat}
 \end{table*}
%

As shown in \fig{tslice_SAD.eps}(c), the early, green-colored contractions have short ranges 
of $10\arcsec$--$20 \arcsec$, because of the compactness of the flare at this time,  
while the later, blue-colored contractions travel great distances 
up to $200 \arcsec= 150 \Mm$. 
However, their statistical medians, modes, and means of the initial velocities 
are quite similar within $200$--$300 \kmps$
(\figs{tslice_SAD.eps}(d) and (e); \tab{table_vstat}), suggestive of
a common origin of these contractions possibly as reconnection outflows.
The initial decelerations of the early contractions are, on the other hand,
about twice higher than those of the later ones, with a median of 2.022 vs.~$0.848 \kmpss$  
(\figs{tslice_SAD.eps}(f) and (g)).

The later, high-altitude contractions have a mode velocity $-158 \kmps$ that is 
similar to the median velocity $-150 \kmps$ of supra-arcade downflows
\citep{SavageS.McKenzieD.SA.downflow.stat.2011ApJ...730...98S, WarrenH.AIA.SADs.2011ApJ...742...92W}.
However, their median and mean velocities, $-242$ and $-282 \kmps$, are nearly twice greater.
This is due to the high velocity tail with 36 out of 174 or 21\% contractions being faster than 
$-400 \kmps$ (hatched area in \fig{tslice_SAD.eps}(e)), reaching a maximum of $-918 \kmps$.
We ascribe this difference to AIA's improved capabilities that allow us to detect
such contractions in their early stages when their velocities are high and emissions are weak.

The initial velocities and decelerations of the earlier and later fast contractions, when taken together,
generally increase with time up to $\sim$06:10~UT and then decrease. This maximum time is delayed by almost 1~h
from that of the slow loop shrinkages and the onset of the HXR burst represented by the \goes flux derivative 
shown in \fig{tslice_shrink.eps}(c). The temporal variations here seem to be
more closely correlated with the \goes flux itself. 
The late-phase decreases of the initial velocities and decelerations
are physically reasonable as energy release subsides,
but there is an observational bias for those tracks starting at the edge of AIA's FOV 
which may have decelerated before they are first detected.

\begin{figure}[thbp]      
 \epsscale{1.1}	
 \plotone{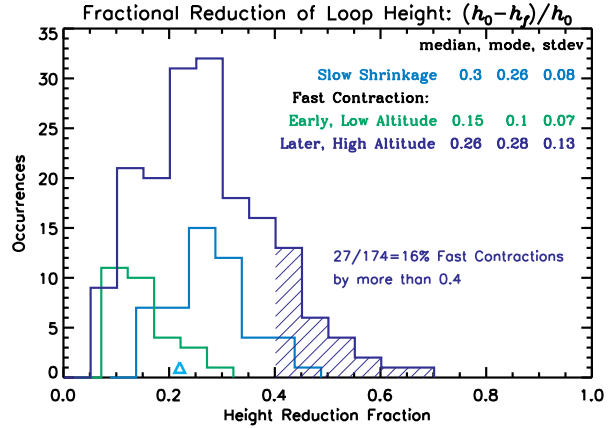}
 \caption[]{\footnotesize
 Histograms of fractional height reductions of slow shrinkages (light blue) and early, low-altitude (green)
 and later, high-altitude (blue) fast contractions. The hatched area indicates 16\% of the later
 fast contractions with reductions greater than 0.4. The cyan triangle marks the descent fraction
 of the 10--16~keV loop-top from \tab{table_cent}.
 } \label{shrink-percent.eps}
 \end{figure}

\fig{shrink-percent.eps} shows the distributions of fractional height reductions of the
downward loop shrinkages and contractions, defined as $(h_0 - h_f)/h_0$, 
where $h_0$ and $h_f$ are the initial and final projected heights. 
The slow shrinkages (light blue) and later, high-altitude fast contractions (blue)
have similar distributions with medians of 0.3 and 0.26, respectively, which are comparable to
those observed in SXRs \citep{ForbesT1996ApJ...459..330F, Reeves.HinodeXRT-shrink.2008ApJ...675..868R}
and predicted theoretically \citep{LinJ.reconn-cusp.2004SoPh..222..115L}.
The early, low-altitude fast contractions (green) have a median of twice smaller,
but comparable to the 10--20\% descent of the overall loop-top emission (\tab{table_cent}). 
Meanwhile, the later fast contractions have a sizable fraction of 16\% (27 out of 174)
with height reductions more than 0.4 (up to 0.7), which is larger than predicted.
Note that these observed values are lower limits, because loops, especially those detected at the edge of AIA's FOV,
could continue contraction before and after their observed intervals when their emission remains below detection.
%

\subsection{Fast Upward Plasmoid Ejections}
\label{subsect_upward}

As shown in \fig{tslice_SAD.eps}, {upward plasmoid ejections} appear as steep, elongated tracks (red fits).
Compared with the downward loop contractions (green fits) at the same time, 
they travel greater distances ($20 \arcsec$--$80\arcsec$ vs.~$10 \arcsec$--$20\arcsec$), 
with larger initial velocities (median: 322 vs.~$-208 \kmps$) 
but smaller decelerations (median: $-0.018$ vs.~$2.022 \kmpss$).
(The opposite signs of the acceleration and velocity here indicate deceleration.)
Some upward ejections even experience acceleration.
Such differences, especially the ratio of their median velocities of $208/322=2/3$, are consistent
with previous observations and simulations of oppositely directed reconnection outflows 
\citep{TakasaoS.Asai.recon.in-outflow.2012ApJ...745L...6T, Barta.plasmoid.up-down-motion.2008A&A...477..649B, 
ShenCC.LinJ.current.sheet.MHD.2011ApJ...737...14S, MurphyNick.asymmetric-reconn.2012ApJ...751...56M,
Karpen.CME-flare-onset.2012ApJ...760...81K}.
These differences are likely due to different environments:
the upflow runs into a region of (partially) open field lines with low plasma density,
while the downflow impinges on high density, closed flare loops that can cause 
strong deceleration.

The starting heights of these upward ejections and their downward counterparts (\fig{tslice_SAD.eps} (c)) 
both increase with time. This suggests an upward development of the reconnection site situated 
between the opposite outflows \citep[e.g.,][]{ShenCC.LinJ.current.sheet.MHD.2011ApJ...737...14S}.
We estimated the height of the reconnection site from a spline fit,  shown as the green dashed line in \figs{tslice_SAD.eps}(a)--(c), 
to the initial heights of the {bi-directional fast outflows}.
Its initial velocity of a few $\kmps$ is similar to that of the initial loop-top ascent, suggesting
the upward development of reconnection being its underlying mechanism.
This estimated reconnection site leaves the AIA FOV at about 06:40~UT.

During 04:15--05:25~UT, we identified 25 upward ejections (red) and 29 downward, early fast contractions (green),
at average occurrence rates of once every 2.4 and 2.2~minutes, respectively.
These rates are comparable to that of the later fast contractions (blue, up to 10:00~UT) of once every 2.0~minutes 
and to those in MHD simulations \citep{Barta.plasmoid.up-down-motion.2008A&A...477..649B}.%
 \footnote{The $\sim$2~minute intervals of fast contractions are comparable to the
 typical periodicities of recently detected quasi-periodic fast-mode magnetosonic wave trains
 \citep{LiuW.FastWave.2011ApJ...736L..13L, LiuW.cavity-oscil.2012ApJ...753...52L,
 Ofman.Liu.fast-wave.2011ApJ...740L..33O, ShenYD.LiuY.QPF.wave.2012ApJ...753...53S}
 that are correlated with flare pulsations. This may indicate that those waves are
 triggered by the energy release episodes associated with these contractions.
 }
The persistence throughout the entire flare of such contractions also agrees with
the statistical result from \yohkoh \citep{KhanJ.SADs-timing.2007A&A...475..333K}.

The initial velocity of upward ejections, as shown in \fig{tslice_SAD.eps}(d), increases with time
and seems to be correlated with the height of the overlying flux rope as it evolves into eruption.
In particular, there is a rapid velocity increase at the onset of the impulsive phase, reaching $1050 \kmps$.
This aspect of this event is independently studied by R.~\citet{LiuR.plasmoid.2013ApJ}.

\section{Spatial Distribution of Energy and Temperature Dependent Emission}
\label{sect_Tmap}

\subsection{\hsi X-ray Spectra}
\label{subsect_spec}

To infer the overall properties of nonthermal particles and thermal plasma,
we analyzed spatially integrated \hsi spectra of individual detectors following the procedures
detailed in \citet{LiuW_2LT.2008ApJ...676..704L} and \citet{MilliganDennis.2009ApJ.EIS-v-T}. 

\begin{figure}[thbp]      
 \epsscale{1.1}	
 \plotone{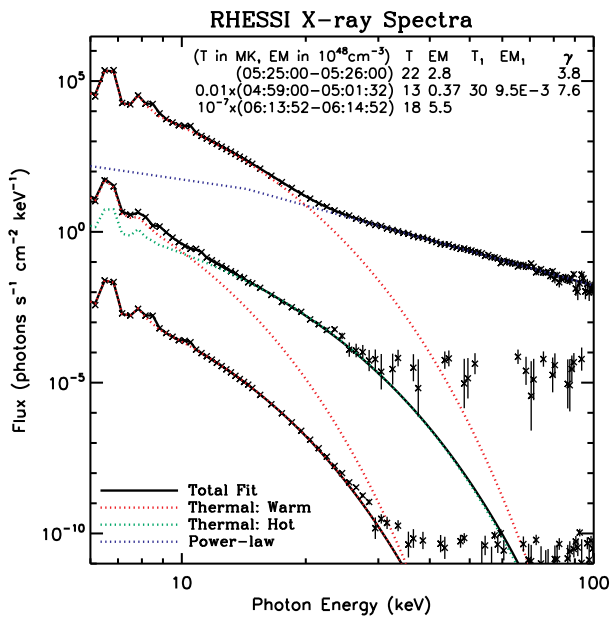}
 \caption[]{\footnotesize	
 \hsi spectra at three selected times before, during, and after
 the impulsive phase. The lower two spectra are vertically shifted by decades to avoid overlap.
 Colored dotted lines are isothermal and power-law fits whose parameters are listed on the top.
 The black solid lines are the total fits.
 } \label{spec.eps}
 \end{figure}

\fig{spec.eps} shows example spectra in three phases of the flare.
 (1) During the pre-impulsive phase (04:59:00--05:01:32~UT), we fitted the spectrum
with two isothermal functions combined: a warm component (red dotted) of temperature $T=13 \MK$
and emission measure ${\rm EM}= 0.37 \E{48} \pcmc$ plus a hot component (green dotted) of $30 \MK$
\citep[also called super-hot,][]{CapisA.LinR.superhotLT.2010ApJ...725L.161C}
and a 40 times smaller EM. If we were to replace the hot component with a power-law $\propto E^{-\gamma}$,
we would have a steep slope of $\gamma= 7.6$.
 (2) During the impulsive phase (05:25--05:26~UT), we used
an isothermal ($T=22 \MK$, ${\rm EM}= 2.8 \E{48} \pcmc$) plus power-law (blue dotted, $\gamma= 3.8$) model.
 (3) During the decay phase,	
the data can be fitted with one isothermal function of a slightly reduced temperature of $18 \MK$ 
but twice higher EM of $5.5 \E{48} \pcmc$. 
Note that at $E$$>$25~keV, the power law in the impulsive phase is more than an order of magnitude higher than
the thermal component. This indicates that the 25--50~keV Masuda-type loop-top source shown in \fig{maps_hsi.eps}
is primarily nonthermal and likely a particle acceleration region itself 
\citep[e.g.,][]{LiuW_2LT.2008ApJ...676..704L, Krucker.LT.acc.all-e.2010ApJ...714.1108K}.
 \begin{figure*}[thbp]      
 \includegraphics[height=2.5in]{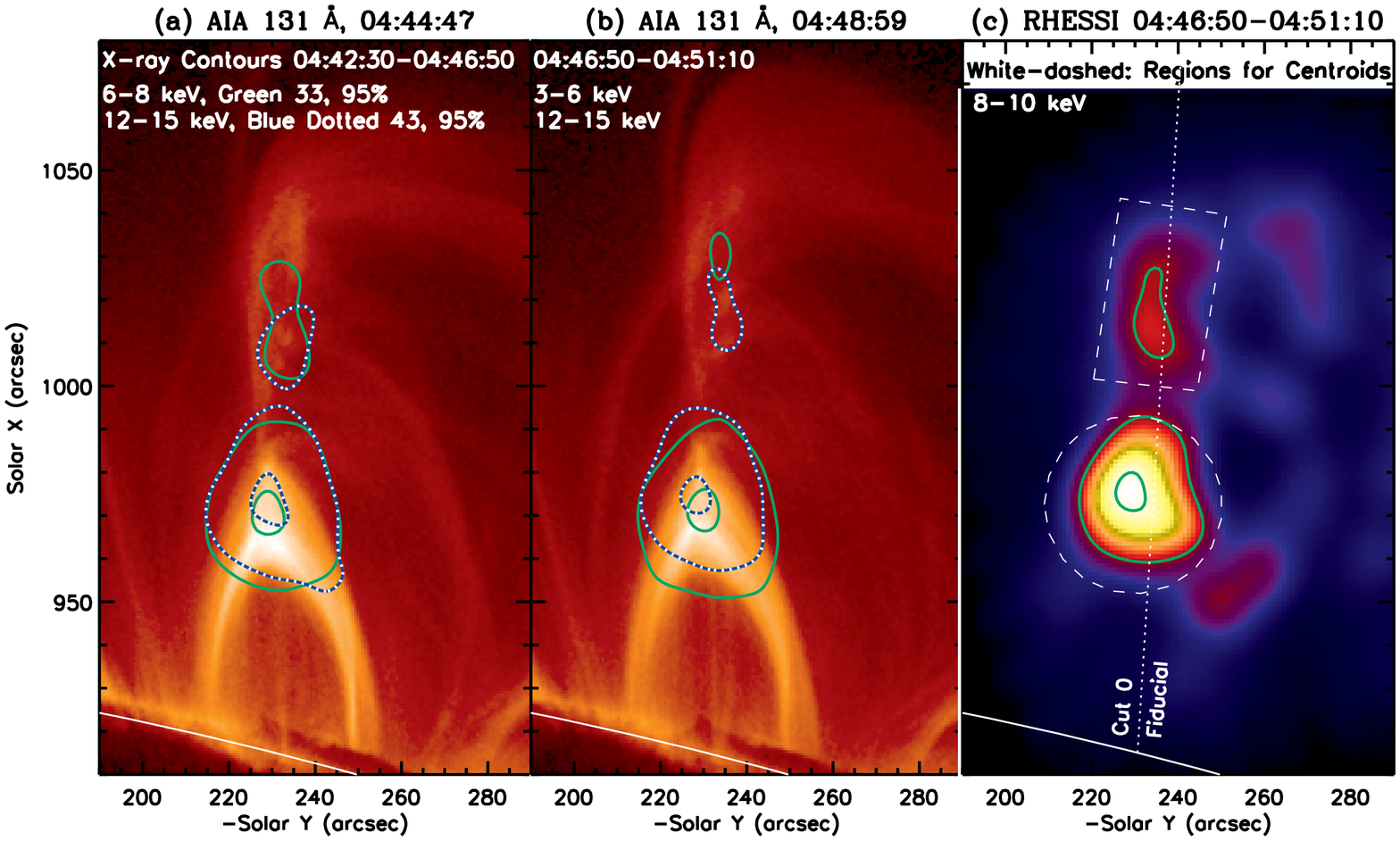}		
 \includegraphics[height=2.5in]{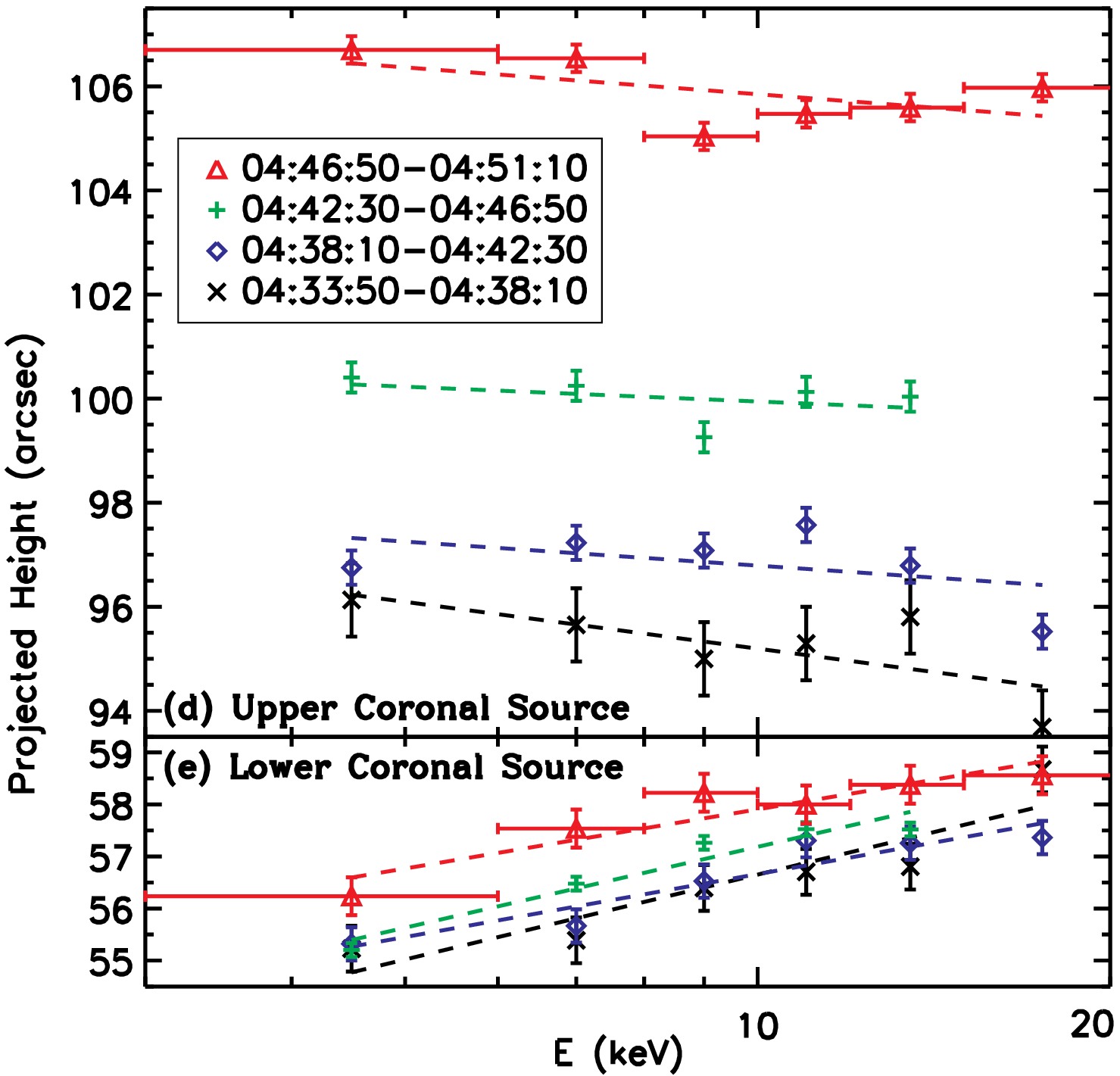}		
 \caption[]{\footnotesize
 Energy dependence of the double coronal X-ray sources likely located in the oppositely directed reconnection outflows.
  (a) and (b) \hsi contours overlaid on AIA 131~\AA\ images.
 The higher energy sources (blue dotted contours at 43\% and 95\% of the image maximum) 
 are closer toward one another than the lower energy sources (green contours). 
  (c) Contoured 6--8~keV image at the time of (b). 	
  (d) and (e) Projected heights along the fiducial Cut~0
 of the centroids of the two sources enclosed by the white dashed lines in (c).
 The result is color-coded for four consecutive times overlaid with linear fits in dashed line.
 The red horizontal error bars indicate energy bins. 
 } \label{cent_E.eps}
 \end{figure*}
%

We applied such spectral fits throughout the flare and
the temporal variations of the fitting parameters are shown in \figs{lc.eps}(d) and (e).
In general, the warm thermal component is persistent in time, and both its $T$ and EM (orange plus signs) increase 
through the HXR burst followed by a more gradual decrease.	
A similar trend is present for the hot thermal component (red crosses) during the pre-impulsive phase. 
A thermal fit to \goes data gives a somewhat lower $T$ but higher EM (black dotted lines), 
because of its well-known preferential response to relatively cooler plasma.
The power-law component (blue) displays a common hardening trend
before the \hsi night data gap.

\subsection{Energy Dependence of Double Coronal Sources}
\label{subsect_cent-E}

In \sect{subsect_ht} we have examined the energy-dependent height distribution
of the loop-top X-ray source (lower coronal source). Here we extend this analysis to the upper coronal source
during the pre-impulsive phase and compare the two sources together.

As shown in \figs{cent_E.eps}(a) and (b), the higher-energy emission (12--15~keV, blue contours) 
of the lower source is located at higher altitudes than lower-energy emission (3--8~keV, green contours), 
while the upper source has an opposite trend.
That is, the two sources are closer to each other at higher energies,
as can be better seen in panel (d) from the linear fits to their centroid heights as a function of logarithmic energy.
This implies higher temperatures of the thermal
plasma and/or harder spectra of the nonthermal particles within the inner regions.
Similar \hsi observations have been reported
\citep{SuiL2003ApJ...596L.251S, LiuW_2LT.2008ApJ...676..704L, LiuW.filmnt.2009ApJ...698..632L}
and interpreted as evidence of magnetic reconnection and associated energy release
being located between the double coronal sources.

At the highest energies ($\gtrsim$10~keV), this trend seems to be reversed,
especially for the upper source. In previous observations (e.g., 
Figure~4 in \citealt{LiuW_2LT.2008ApJ...676..704L}),
more obvious reversals occurred at slightly higher energies of 15--20~keV 
and were interpreted as a transition from the low-energy thermal
regime to the high-energy nonthermal regime, in which greater stopping distances of higher energy electrons
can cause higher energy bremsstrahlung X-rays being located farther away. 
We further suggest that this reversal may be an indication of the two X-ray sources
being {\it spatially separated}, instead of merging together at the highest energies;
so are the locations of primary heating and particle acceleration
from magnetic reconnection.
This is supported by the {\it separate temperature peaks} 
at the two sources as shown in \figs{tem_map.eps}(a) and (h) and discussed in \sect{subsect_AIA_Tmap}.

\subsection{AIA Temperature Maps}
\label{subsect_AIA_Tmap}

 \begin{figure*}[thbp]      
 \includegraphics[height=5.4in]{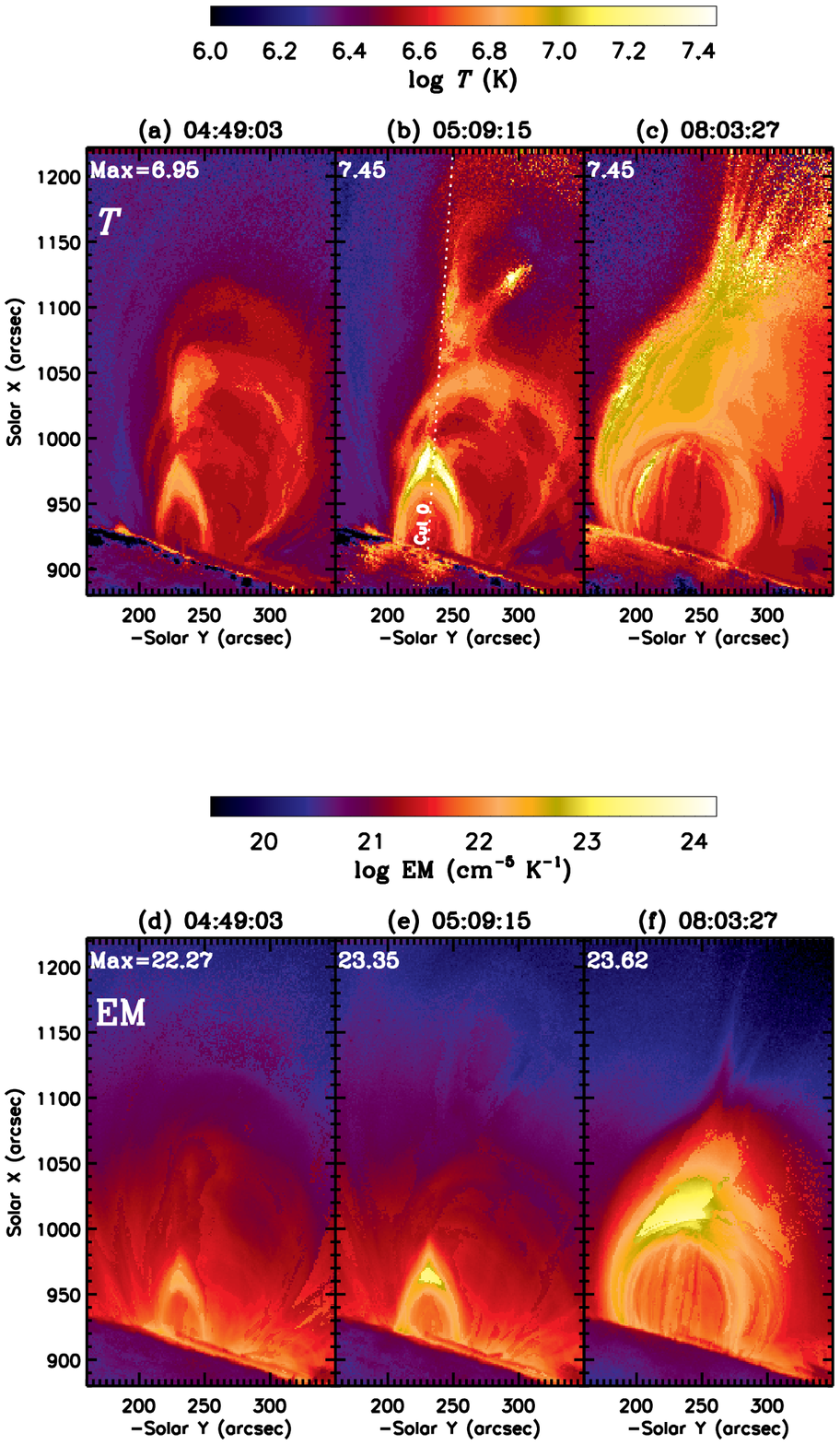}	
 \includegraphics[height=5.4in]{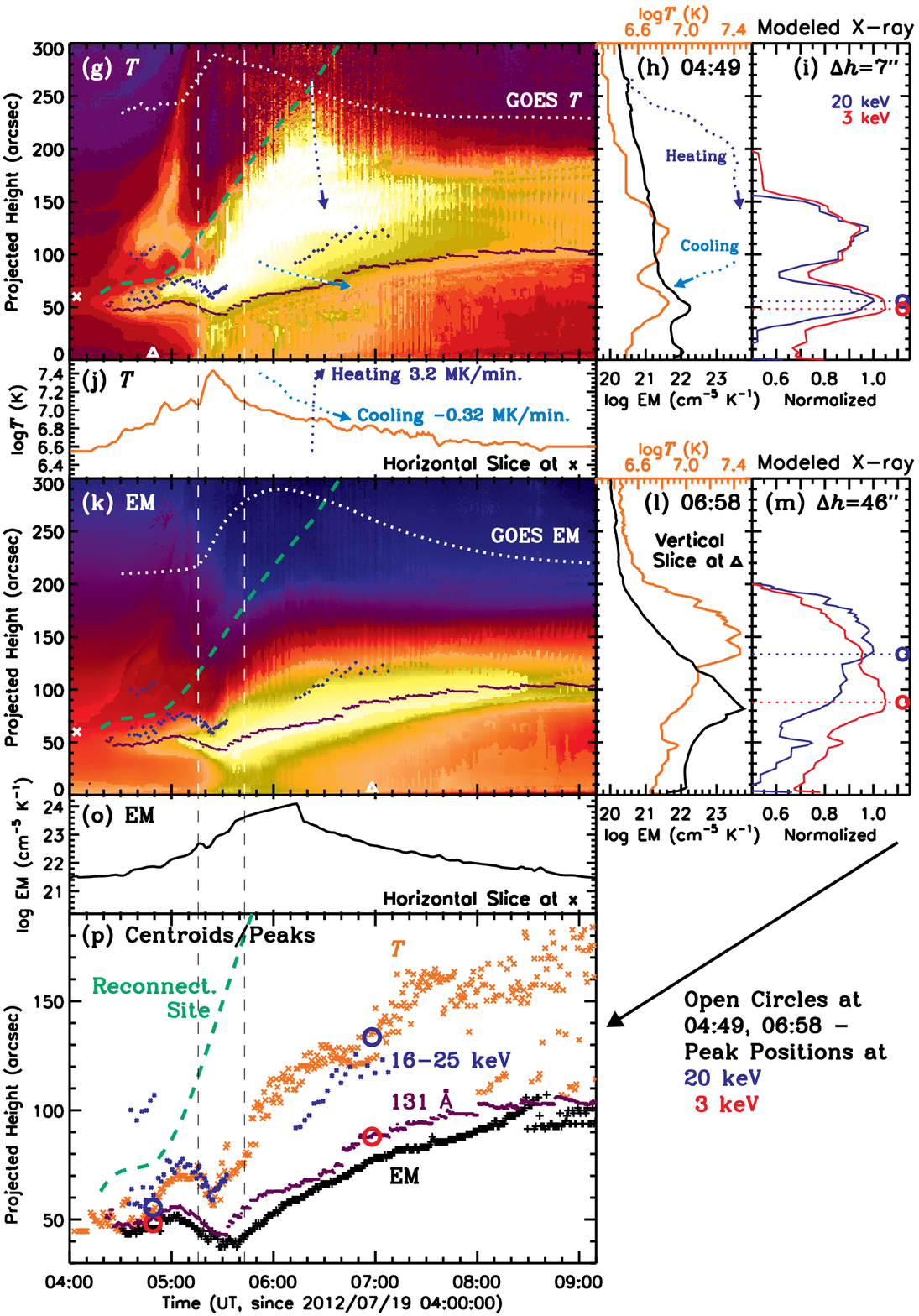}	
 \caption[]{\footnotesize
   Maps of temperature $T$ (a--c) and emission measure EM (d--f) 
 inferred from AIA filter ratios at three selected times on logarithmic scales. 
  (g) and (k) Space-time plots of $T$ and EM maps from Cut~0
 as shown in (b). 	
 The white dotted lines are temperature and emission measure from \goes data 
 as shown in \figs{lc.eps}(d) and (e).
   The first side panels ((h) and (l)) on the right show the height distributions of 
 $T$ (red, top scale) and EM (black, bottom scale) 	
 at the time marked by the triangle on the left.
   The second side panels ((i) and (m)) show the modeled corresponding thermal X-ray emission
 at two photon energies, 3 and 20~keV. 	
 The open circles mark the heights of the emission peaks that are also shown in (p).
   (j) and (o) Temporal profiles $T$ and EM, obtained from a horizontal
 slice of the corresponding space-time plots above at
 height $h_{\rm ref}=60\arcsec$ marked by the cross.
 The arrowed blue and light blue dotted lines in (g) are selected tracks of 
 a fast contraction and slow shrinkage from \figs{tslice_SAD.eps} and \ref{tslice_shrink.eps}.
 The temperatures sampled by these tracks are shown as a function of time and distance in (j) and (h),
 respectively, indicating heating and cooling experienced by these loops as they travel.
   (p) Projected heights of the maxima of $T$ (orange) and EM (black) at each time
 in the corresponding space-time plots.
 Overlaid here and in the above space-time plots are the heights of the 16-25~keV (blue) 
 and 131~\AA\ (purple) loop-top centroids in small symbols from \fig{cent_time.eps}(e),
 and of the inferred reconnection site in green dashed line from \fig{tslice_SAD.eps}(c). 
 The vertical dashed lines mark the impulsive phase.
 } \label{tem_map.eps}
 \end{figure*}
%
To infer the spatial distributions of temperature $T$ and emission measure (EM),
we employed a forward-fitting algorithm using AIA filter ratios
(\citealt{Aschwanden.auto.AIA-T-EM.2011SoPh..tmp..384A}; for a different approach, see
\citealt{Battaglia.Kontar.AIA-DEM.2012ApJ...760..142B}).
This algorithm assumes a Gaussian differential emission measure that has a peak EM at temperature $T$.
Such a model provides the simplest description of the EM weighted temperature distribution 
along the line of sight.


\fig{tem_map.eps} (left) shows examples of $T$ and EM maps.
Early in the event at 04:49~UT,
the highest temperature of $\log T = 6.95$ or $T = 8.9 \MK$ is located in the 
rising flux rope and the underlying flare cusp, but EM is highly concentrated in the latter. 
The temperature then increases with time especially in the flare cusp, 
reaching $\log T = 7.45$ or $T = 28.2 \MK$ at 05:09~UT,
close to the 35~MK of the hot \hsi component at this time (see \fig{lc.eps}(d), red crosses).
In the decay phase, the highest temperatures are located in the outer layer of the flare arcade 
and the lower portion of the contracting cusps (ray-like features) above it.
The EM in the latter is nearly four orders of magnitude lower than in the central arcade,
similar to those of hot fans of rays observed by \yohkoh \citep{SvestkaZ.SXT-hot-fan.1998SoPh..182..179S}.
The considerable increase of EM and thus density in the flare loops from panel (d) to (e), 
prior to the impulsive phase, suggests chromospheric evaporation driven
by thermal conduction \citep{LiuW.HDparticle-I.2009ApJ...702.1553L, Battaglia.conduct-evapor_2009A&A}
or by \Alfven waves \citep{HaerendelG.chromosph.evapor.by.Alfven.wave.2009ApJ...707..903H}, 
rather than electron beam heating that may be important later in the presence of HXR footpoints.


To better follow the history of $T$ and EM, we obtained their space-time plots, 
as shown in \figs{tem_map.eps}(g) and (k), from corresponding maps using Cut~0.
We identified the maximum temperature and EM at each time 
and show their projected heights as orange and black symbols in panel~(p).
In general, both peaks follow the same upward-downward-upward motions as the loop-top emission centroids.
The temperature peaks are near the \hsi 16--25~keV centroids (blue),
while the EM peaks are close to or slightly lower than the AIA 131 \AA\ centroids (purple)
at lower heights. 



The {\it peak offset} of the temperature and EM can be better seen in their height distributions
at selected times	
shown in \figs{tem_map.eps}(h) and (l). It explains the observed energy dispersion 
of the X-ray loop-top centroids shown in \fig{cent_time.eps}(e)
because of the exponential shape of thermal bremsstrahlung spectra (equation~(\ref{thBremeq})).
As shown by the green and red lines in the middle of \fig{spec.eps},
a higher temperature but lower EM produces a harder (shallower) spectrum of lower normalization
that dominates at high energies,
while a lower temperature but higher EM produces a softer (steeper) spectrum of higher normalization
that dominates at low energies.
The temperature peak being located above the EM peak
thus shifts the higher energy X-rays toward greater heights.
Otherwise, if the temperature and EM peaks are cospatial, they would dominate X-rays
at all energies and there would be no separation of centroids with energy.
We note in \figs{tem_map.eps}(h) that the EM decreases more gradually near the flux rope. 
This may lead to the less pronounced energy dispersion of the upper coronal source there
than the lower (loop-top) source (see \fig{cent_E.eps}).

As a proof of concept, we modeled the observed X-rays of energy $E$
with thermal bremsstrahlung radiation	
\citep[][p.~114]{Tandberg-Hanssen.Emslie.1988psf..book.....T}:
 \beq  I_{\rm SXR} \propto ({\rm EM}) {\exp(-E/kT) \over E \sqrt{T} } g(E/kT) , 
 \label{thBremeq}\eeq      	
where	
${\rm EM}$ is the emission measure 	
and $g(E/kT) = (kT/E)^{2/5} $ is the Gaunt factor.
The resulting X-ray profiles at two selected times	
are shown in \figs{tem_map.eps}(i) and (m) for the corresponding $T$ and EM profiles.
As expected, the higher energy 20~keV emission (blue) is dominated by
the temperature peak at a higher altitude, while the 3~keV emission (red)
is dominated by the EM peak at a lower altitude.
Their emission peaks, marked by open circles, differ in height by 
$\Delta h = 7 \arcsec$ and $46 \arcsec$ for the two times. Their peak heights
are repeated in panel~(p) and are close to those of observed loop-top centroids.


As shown in \figs{tem_map.eps}(g) and (p), the high temperature region and particularly the temperature peak
are close to the X-ray loop-top centroids but always below the reconnection site 
(green dashed line). This indicates that primary plasma heating
takes place in reconnection outflows. 	
To illustrate this, we can follow a contracting loop and the temperature variation it senses
as it travels away from the reconnection site. A selected track from \fig{tslice_SAD.eps}(c)
is shown here as the blue dotted arrow. The temperature history on its path, as shown
in \figs{tem_map.eps}(h) and (j), indicates rapid heating
at an average rate of $3.2 \MK \, {\rm min}^{-1}$ from 3.5 to 28~MK over 8~minutes
and $\Delta h = 90 \Mm$. This example gives us a sense of the heating rate averaged along the line of sight,
not necessarily of a specific loop.

Likewise, the track of a slow loop shrinkage at lower altitudes reveals
cooling at an average rate of $-0.32 \MK \, {\rm min}^{-1}$ from 25 to 7.9~MK within 54~minutes.
The rate is $-1 \MK \, {\rm min}^{-1}$ earlier during the impulsive phase in the 28--9~MK range
and $-0.1 \MK \, {\rm min}^{-1}$ later during 07:00--08:30~UT in the 14--7~MK range.
These cooling rates are somewhat lower than previously found at even lower temperatures 
\citep{Vrsnak.2003Nov03shrink.2006SoPh..234..273V}.

\section{Conclusion}    
\label{sect_conclude}


\subsection{Summary}		
\label{subsect_summary}

We have presented detailed EUV and X-ray observations of the 2012 July 19 M7.7 flare,
focusing on the signatures of {\it magnetic reconnection} 	
and associated energy release. We summarize our findings and discuss their implications as follows.

\begin{enumerate}	

\item	

The V-shaped EUV emission on the trailing edge of the flux rope CME and the underlying, 
inverted V-shaped flare loops, both associated with distinct X-ray emission, 
suggest two oppositely oriented Y-type null points with a vertical {\it current sheet}
formed in between (\figs{maps_overview.eps} and \ref{maps_blob.eps}). 

\item	

Originating from the inferred magnetic reconnection site within the current sheet 
are { {\it bi-directional outflows}} in the forms of plasmoids and cusp-shaped loops. 
The upward ejections have a median initial velocity of $320 \kmps$ and a maximum of $1050 \kmps$,
while the concurrent downward cusp contractions are $\sim$$2/3$ slower with a median of $-210 \kmps$.   
Even faster contractions up to $-920 \kmps$   
occur after the flux rope eruption for another 10~h,
and a sizable fraction of 21\% of them have speeds $\geq$$400 \kmps$, 
twice faster than previously reported
\citep{SavageS.McKenzieD.SA.downflow.stat.2011ApJ...730...98S}.	
Such high velocities are comparable to expected coronal \Alfven speeds of $\sim$$1000 \kmps$ 
and sound speeds of 520--$910 \kmps$ for a 10--30~MK flaring plasma.
At lower altitudes, flare loops persistently shrink at typical initial velocity of $-17 \kmps$
and gradually decelerate to about $-5 \kmps$ within 0.5--2.5~h.
They are all evidence of {\it reconnection outflows} during different stages (\figs{tslice_shrink.eps} and \ref{tslice_SAD.eps}).

\item	

The {\it double coronal X-ray sources}        
are spatially separated and located in the regions of { bi-directional plasma outflows}.	
The highest temperature is found near the loop-top X-ray source well below the reconnection site.
This suggests that primary plasma heating and particle acceleration take place in
the {\it reconnection outflow regions} \citep{Holman.flare.review.2012PhT....65d..56H}, 
rather than at the reconnection site itself. Models with this ingredient
were proposed long ago \citep[e.g.,][]{ForbesT.PriestE.flare-LT-shock.1983SoPh...84..169F},
but solid observational evidence as presented here has been lacking (\figs{cent_E.eps} and \ref{tem_map.eps}).

\item   

An {\it energy dispersion} is present in the loop-top position with higher-energy X-ray
and hotter EUV emission located at greater heights over a large range of 
$\Delta h \geq 26 \Mm$. The 25--50~keV nonthermal emission lies $15 \Mm$ 
(twice that of the Masuda flare) above the 6--10~keV thermal emission. 
This agrees with the expected trend of softer electron spectra in the nonthermal regime
and lower temperatures in the thermal regime being associated with earlier energized loops, 
which are located further below the primary locus of energy release.
The upper coronal X-ray source    
has an opposite trend because it is located in the oppositely directed reconnection outflow 
(\figs{maps_hsi.eps}, \ref{cent_time.eps}, and \ref{cent_E.eps}).  

\item	

Prior to the recently recognized descent followed by a continuous ascent, 
the overall loop-top emission    
experiences an {\it initial ascent} for nearly an hour.
This is the first time that such motions, including the descent, are observed simultaneously from EUV to HXRs
covering a wide range of temperatures of 1--30~MK.   
The transition from ascent to descent coincides with the rapid acceleration of the flux rope CME.
The loop-top descends at $\sim$$10 \kmps$, about 50\% slower in EUV than in X-ray
(\fig{cent_time.eps}).   

\item	

The flare impulsive phase starts when a rapid velocity increases occur for
the overall loop-top descent, the individual loop shrinkages, and the upward plasmoid ejections. 
This is delayed by 10 minutes from the initial loop-top descent, implying that the energy release rate
is more intimately correlated with these velocities than the loop-top position (\fig{cent_time.eps}).

\end{enumerate}		

\subsection{Proposed Physical Picture}	
\label{subsect_model}

\begin{figure*}[thbp]      
 \epsscale{0.35}	
 \plotone{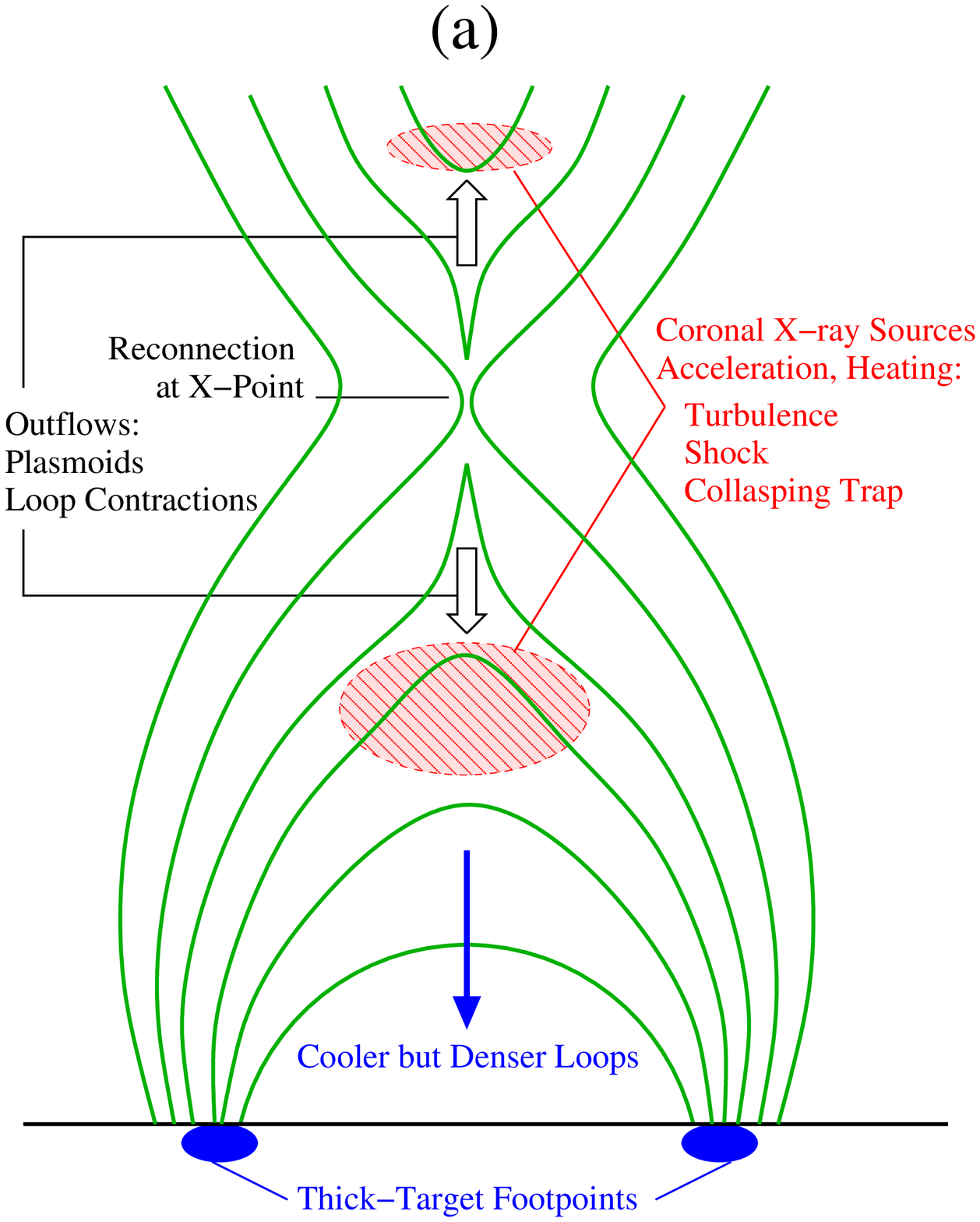}
 \epsscale{0.14}	
 \plotone{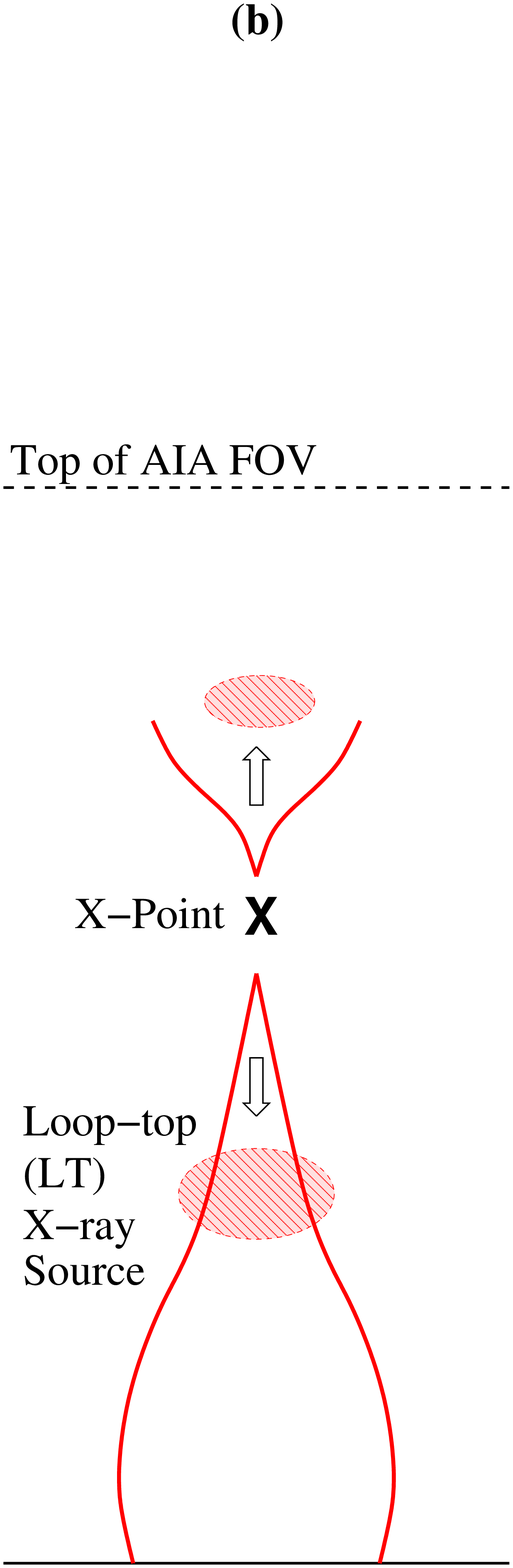}
 \epsscale{0.14}	
 \plotone{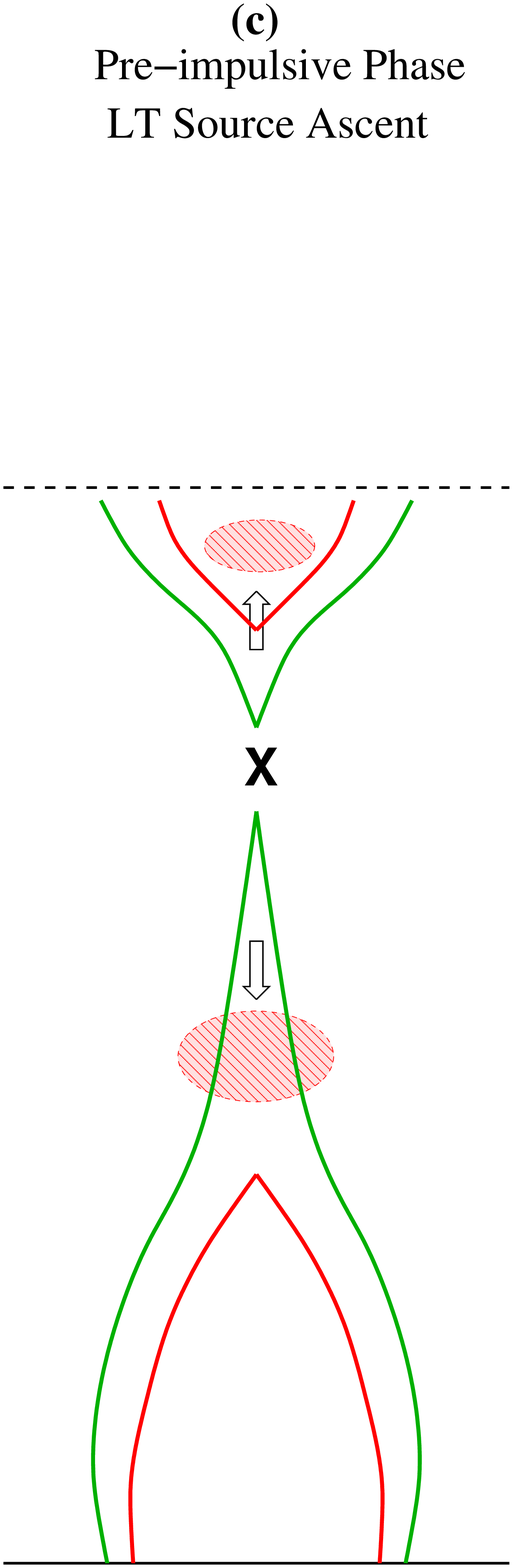}
 \epsscale{0.14}	
 \plotone{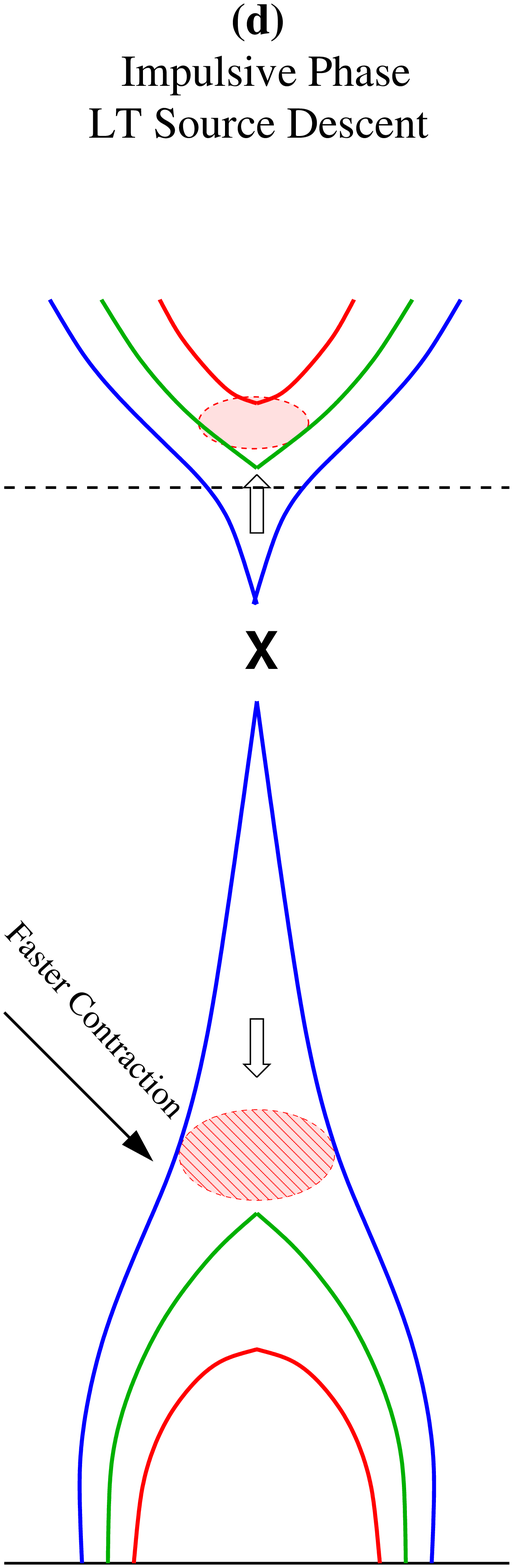}
 \epsscale{0.14}	
 \plotone{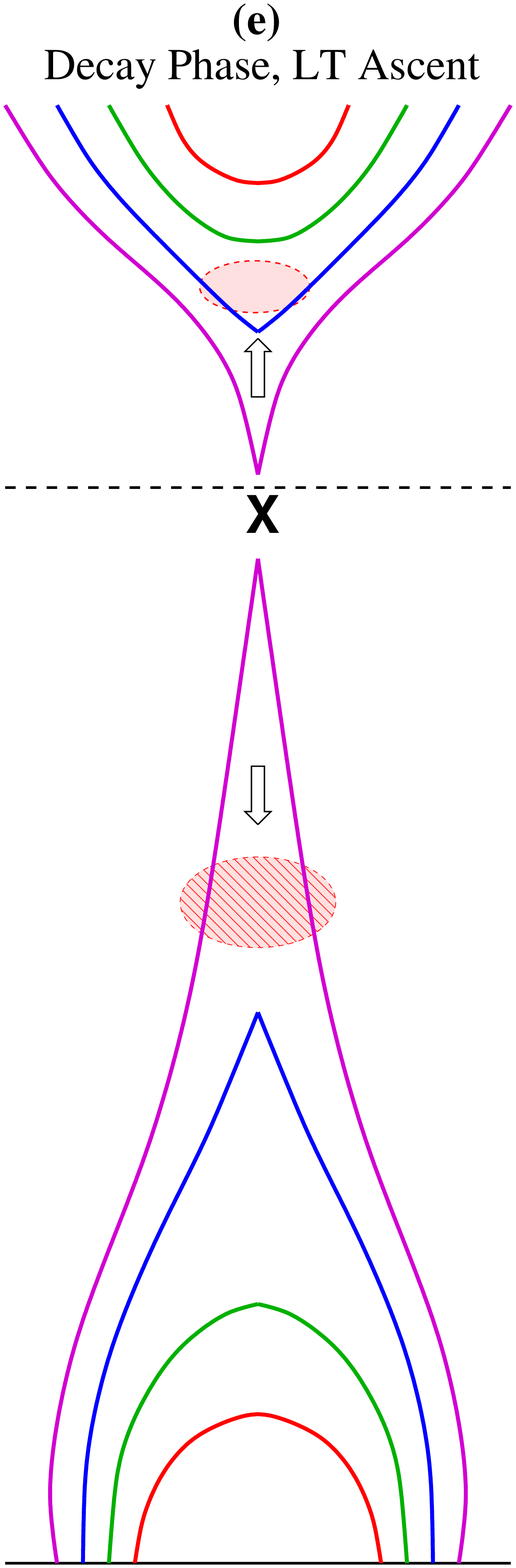}		
 \caption[]{\footnotesize
 (a) Schematic of the proposed flare model in which particle acceleration and plasma heating
 take place in the reconnection outflow regions away from the reconnection site.
 (b)--(e) Temporal development of reconnection and the up-down-up three-stage motion 
 of the loop-top (LT) X-ray source. 
 { We show only the portions of reconnected magnetic field lines near 
 the reconnection site, above and below which each pair of field lines are identified by 
 the same color during their relaxations.	
 Note that the higher contraction speeds (e.g., of the green loop) from (c) to (d) is associated 
 with the loop-top source descent.
 The upper coronal source is not hatched during the impulsive and decay phases,
 indicating its non-detection due to weak emission.
 We assume that bi-directional reconnection outflows and relaxations of field lines 
 persist throughout the event, but the upward component is beyond the AIA FOV
 during the late phase and thus not observed.
 }} \label{model.eps}
 \end{figure*}

We propose the following physical picture to tie together the observations.
{ The major points, including the location of particle acceleration
and the three-stage (up-down-up) motion of the loop-top X-ray source, are sketched in \fig{model.eps}.}
%

The earlier {\it confined} C4.5 flare leads to the formation of a flux rope and cusp-shaped
loops underneath it. After six hours of slow evolution, 
the flux rope becomes unstable and rises
\citep{Patsourakos.pre-exist-flux-rope_CME.2013ApJ...764..125P}. 	
A vertical current sheet forms between two Y-type null points
at the lower tip of the flux rope and the upper tip of the underlying cusp.
Magnetic reconnection ensues within the current sheet, 
leading to the {\it eruptive} M7.7 flare.

Magnetic reconnection  produces {\it bi-directional outflows} 
in forms of the observed plasmoids and contracting loops.
Plasmoids are flux tubes formed by the tearing mode \citep{FurthH.tearing-mode.1963PhFl....6..459F}
with a guiding field along the current sheet.
They are magnetic islands in two dimension or when the current sheet is seen edge-on.
These outflows are driven by the magnetic tension force of the highly bent,
newly reconnected field lines, as in those pointed cusps. 
The outflows generally decelerate, as observed here, when they run into the ambient corona
and when the contracting loops relax to less bent shapes with reduced magnetic tension.
This implies that the low-altitude slow shrinkages could be
the late stages of decelerated high-altitude fast contractions.	

{
Several mechanisms can operate in the reconnection outflows and contribute to
particle acceleration and plasma heating.
{\it Turbulence or plasma waves}, for example, can be generated by the
interaction of the high speed flows with the ambient corona.
Upon cascading to smaller scales (comparable to the gyro-radii of background particles)
at some distance from the reconnection site,
turbulence can accelerate the particles and heat the plasma 
\citep{HamiltonR1992ApJ...398..350H, MillerJ1996ApJ...461..445M,
ChandranB.slow-mode.acc.flare.2003ApJ...599.1426C,
PetrosianV2004ApJ...610..550P, 		
PetrosianV.Yan.Lazarian.damping.2006ApJ...644..603P, 
JiangY.cascade.damping.2009ApJ...698..163J, FleishmanGD.SA-acc-helical-turb.2013MNRAS.tmp..447F}.
Additional contribution to particle acceleration and/or heating can come from 
fast-mode {\it shocks} in the outflow regions	
\citep{ForbesT.PriestE.flare-LT-shock.1983SoPh...84..169F, 
TsunetaS1998ApJ...495L..67T, GuoF.Giacalone.shock-acc-flare.2012ApJ...753...28G, 
Nishizuka.Shibata.Fermi-shock-acc.2013arXiv1301.6242N},
gas dynamic shocks within contracting flux tubes
\citep{LongcopeD.gas-shock-retracting-loop.2009ApJ...690L..18L},		
and the {\it first-order Fermi} and {\it betatron} mechanisms
within the collapsing traps formed by contracting loops
\citep{Somov.Kosugi.collps.trap.1997ApJ...485..859S, Karlicky.Kosugi.collps-trap.2004A&A...419.1159K,
KarlickyM.collaps-trap.2006SSRv..122..161K, GradyK.collaps.trap.acc.2012A&A...546A..85G}.

}

As the flux rope evolves into its fast rise and eruption stage, the current sheet could 
become sufficiently long causing significant tearing instability 
\citep[e.g.,][]{Barta.plasmoid.up-down-motion.2008A&A...477..649B},
and/or become thinner than the ion skin depth leading to a transition 
from Sweet-Parker reconnection to collisionless Hall reconnection
\citep{CassakP.fast.reconnection.2006ApJ...644L.145C}.		
Both can result in enhanced rates of reconnection and energy release.
The outflow velocity will also increase, producing stronger heating and particle acceleration 
by one or a combination of the above three mechanisms (turbulence, shock, and betatron)
whose efficiency is expected to be positively correlated with the outflow velocity. 
For example, faster outflows can generate stronger turbulence that can
lead to stronger particle acceleration \citep[see][for a comparison between acceleration 
by turbulence and shocks]{PetrosianV.SA.review.2012SSRv..173..535P}.
This explains the observed temporal correlation of the 
increase in outflow velocity and the onset of the impulse phase and HXR burst.


The loop-top source seen in X-ray and EUV represents the collective emission
of the ensemble of flare loops, each of which undergoes post-reconnection contraction or shrinkage. 
The loop-top emission centroid is the average position of this ensemble, which in the thermal regime is
determined by the convolution of the spatial distributions of temperature and emission measure.
{ 
Such distributions are influenced by several competing processes, including energization of 
new loops produced by the {\it upward} developing reconnection,
and {\it downward} contractions and cooling of previously reconnected loops.
We suggest that the interplay of these processes results in the 
up-down-up migration of the loop-top source.
Around the early impulsive phase, the observed higher velocities of loop contractions 
tend to shift the overall loop-top source downward, as depicted in \figs{model.eps}(c) and (d). 
Within the contracting loops, 
conductive cooling can be considerably reduced by stronger turbulence produced by 
faster reconnection outflows \citep{ChandranB1998PhRvL..80.3077C, JiangY2006ApJ...638.1140J}.
Slower cooling helps the X-ray/EUV emission of these hot loops last longer during their contractions
and further contribute to the loop-top descent.
Before and after the early impulsive phase, the situation could be different.
Loop contractions are slower and associated faster cooling can make 
these loops cool below the instrumental temperature passband more rapidly.	
The upward development of reconnection could thus dominate,
leading to the upward loop-top migration.
This interpretation is supported by the observed temporal coincidence of 
the highest velocities of individual loop shrinkages and of the loop-top descent 
at the impulsive phase onset.
Like cooling, {\it chromospheric evaporation} can also play a role.	
After the initial chromospheric evaporation associated with energization of newly reconnected loops, 
if it continues operating during the subsequent loop contractions,
the densities and thus emission measures of these loops would keep rising
and contribute to the downward loop-top drift.
Otherwise, if it rapidly subsides, it would assist the upward loop-top drift.
Investigating such a complex interplay would require detailed numerical modeling,
which is beyond our scope here.
}



\subsection{Discussion}
\label{subsect_discuss}

The fast downward contractions of EUV loops are best seen during the decay phase, likely because of their greater distances traveled
making them easier to be detected.
Likewise, different environments may explain the factor of two difference in the median deceleration
between the early and later fast contractions (\figs{tslice_SAD.eps}(f) and (g)).
The earlier, compact flare loops of stronger magnetic field can produce stronger resistance
\citep[e.g., by the Lorentz force from a reverse current;][]{Barta.plasmoid.up-down-motion.2008A&A...477..649B}
to quickly brake the contracting loops impinging on them from above,    
while the later, large flare arcade of weaker field at greater heights
and greater travel distances 	
may allow more gradual deceleration.

The X-ray spectra are essentially thermal during the decay phase. 
This suggests that the late phase contractions are associated with plasma heating 
rather than particle acceleration. This is expected because the magnetic field strength
decreases with height, and so does the available magnetic energy to be released by reconnection. 
Such prolonged heating 	
may contribute to the well-known slower than expected cooling of flare plasma 
(see \fig{lc.eps}(d); \citealt{McTiernanJ.Yohkok.SXT.decay.1993ApJ...416L..91M, JiangY2006ApJ...638.1140J}).

The fast high-altitude contractions and slow low-altitude shrinkages
appear as two statistically distinct populations. 
The median {\it final} velocity of the former is more than four times 
the median {\it initial} velocity of the latter ($-73$ vs.~$-17 \kmps$; \figs{tslice_shrink.eps} and \ref{tslice_SAD.eps}).
We have interpreted the latter as the late stages of the former that have decelerated,   
but we rarely see a continuous track decelerating from $>$$400 \kmps$ to a few $\kmps$.
Alternatively, this distinction may be due to inhomogeneity of reconnection
\citep[see, e.g.,][]{AsaiA.SADs.2002-07-23-flare.2004ApJ...605L..77A}
in the current sheet above the flare arcade that is primarily oriented along the line of sight.
The {\it persistent} slow shrinkages could result from reconnection at a moderate rate throughout
the current sheet, which produces flare loops forming the arcade and double ribbons
at their footpoints. The {\it episodic} fast contractions could be signatures
of plasmoids within the current sheet, associated with enhanced reconnection and energy release rates. 
These fast contracting loops are dispersed along the line of sight amidst slowly shrinking loops, 
producing the observed effects (e.g., \fig{tslice_SAD.eps}(b)). 


There are other physically different (though not necessarily unrelated) 
phenomena that share similar observational signatures
and should not be confused with the contractions of {\it newly reconnected} loops studied here. 
For example, {\it pre-existing} coronal loops can contract at typical speeds $<$$100 \kmps$    
during eruptions \citep{LiuRui.fast.loop.contraction.2010ApJ...714L..41L, LiuR.contract.loops.2012ApJ...757..150L, 
SunXD.HMI.2011Feb15.X1.5.implosion.2012ApJ...748...77S}, 
interpreted as implosion \citep{HudsonH.implosion.2000ApJ...531L..75H, Janse.Low.implosion.2007A&A...472..957J} 
due to the rapid release of magnetic energy and thus reduction of magnetic pressure within the eruption volume. 
When a CME eruption initiates at an elevated height, it can produces a downward push to displace ambient coronal loops
\citep[e.g.,][see their Figure 12(h), at $-60 \kmps$]{LiuW.cavity-oscil.2012ApJ...753...52L}.
Even slower shrinkage at $-3 \kmps$ can occur in active region loops due to gradual cooling
\citep{WangJX.loop-shrink.1997ApJ...478L..41W}.
Sometimes, conjugate X-ray footpoints approach each other while the loop-top source descends
\citep{JiH.converge.2006ApJ...636L.173J,   
LiuW_FPAsym_2009ApJ...693..847L, YangYH.FPstat.2009ApJ...693..132Y},    
likely because of reconnection progressing toward less sheared loops at lower altitudes.
We have not found these signatures in the flare under study here, 
except for a few overlying existing loops that
contract at about $-150 \kmps$ during 05:40--05:50~UT near the end of the impulsive phase.
In addition, some high-lying loops continue to expand and erupt at $\sim$$100 \kmps$ 
up to 08:00~UT, three hours after the original flux rope eruption.
Examination of such features and quantitative comparison of these observations with 
particle acceleration models will be subjects of future investigations.


\acknowledgments
{W.L. was supported by NASA \sdoA/AIA contract NNG04EA00C and LWS/TR\&T contract NNX11AO68G. 
Q.R.C. and V.P. were supported by NASA grants NNX10AC06G and NNX13AF79G.
W.L. thanks Marian Karlicky, Paul Cassak, Hugh Hudson, Sasha Kosovichev, 
Judy Karpen, Rick DeVore, Greg Slater, Tom Berger, and Mark Cheung for useful discussions.

} 

\section*{Appendix A \\
 Observations of All AIA Channels}
\label{append_sect:slices}

\begin{figure*}[thbp]      
 \epsscale{0.37}	
 \plotone{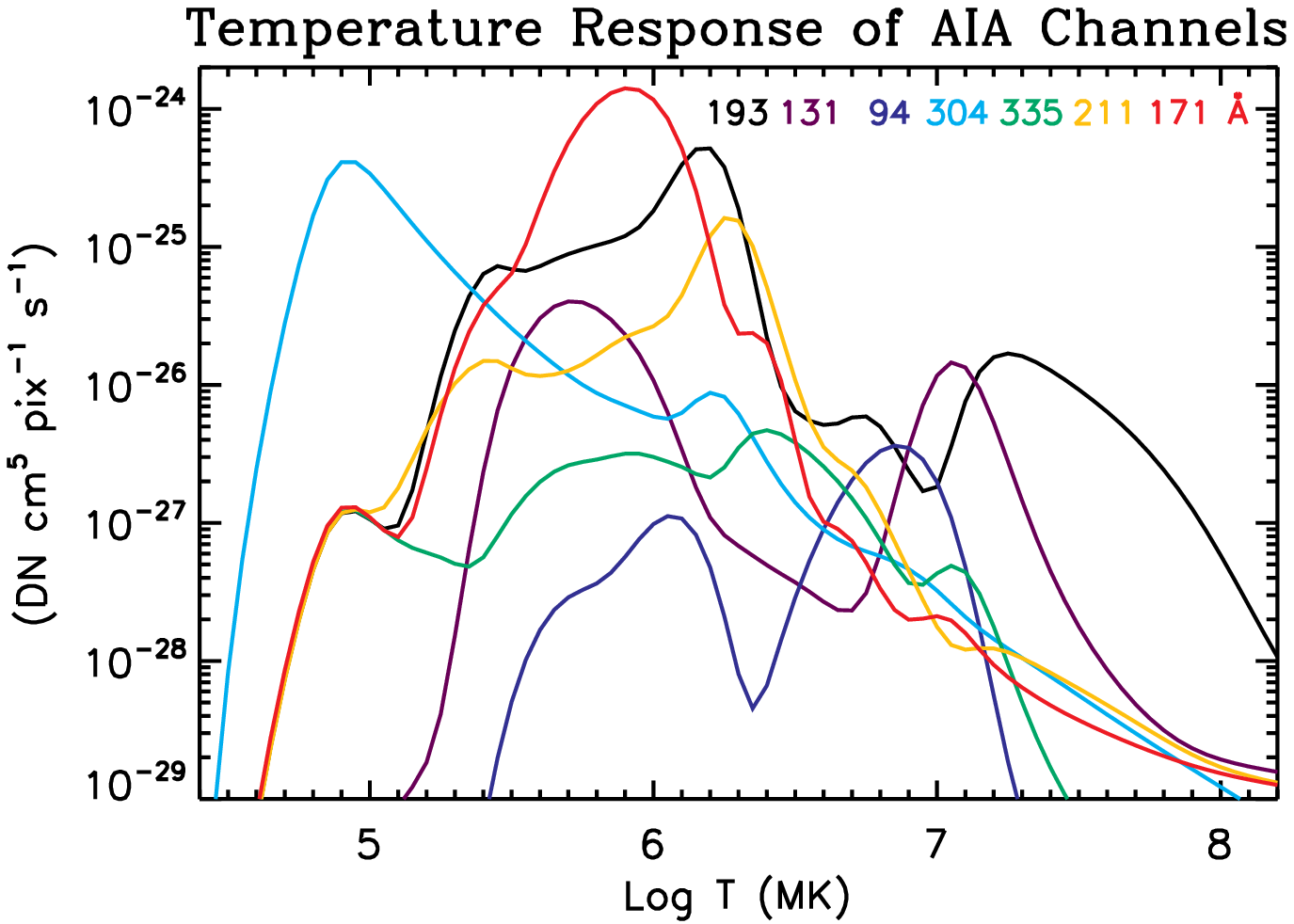}
 \epsscale{0.83}	
 \plotone{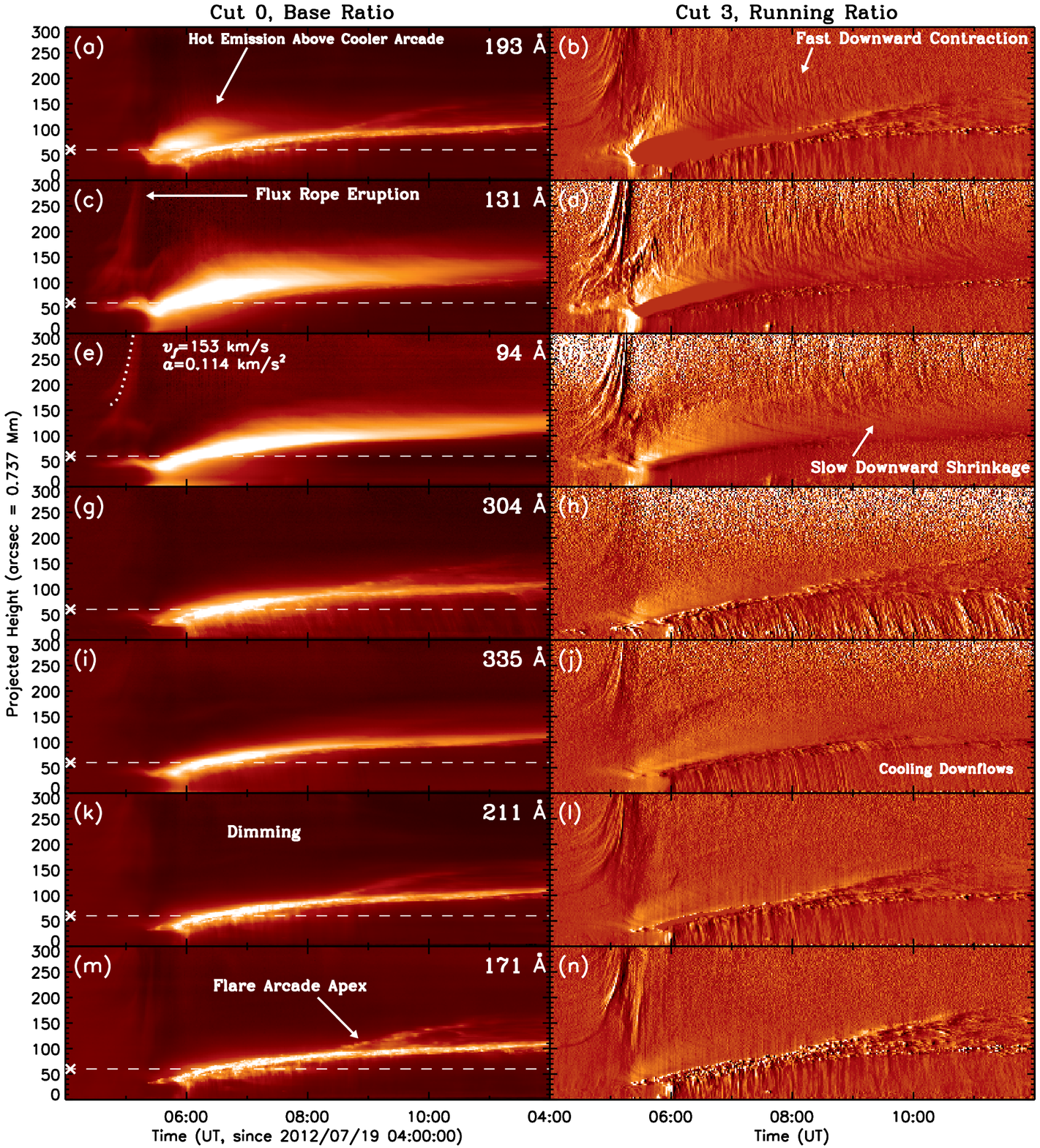}		
 \caption[]{\footnotesize
 Top: Temperature response of AIA EUV channels.
 Other panels: Space-time plots of all AIA EUV channels in base ratio from Cut~0 on the left
 and in running ratio from Cut~3 on the right.  The panels are arranged approximately
 in the order from high to low temperature response.
 The cross and horizontal dash line on the left mark the slice for obtaining the temporal
 profiles shown in \fig{lc.eps}(b). The dotted line in (e) is a parabolic fit to 
 the central portion of the erupting flux rope CME that has an average acceleration
 of $a= 0.114 \pm 0.002 \kmpss$ and achieves a velocity of $v_{f}= 153 \pm 3 \kmps$
 at the edge of AIA's FOV.
 } \label{tslice_all.eps}
 \end{figure*}

In the main text, we focused on the 131~\AA\ channel. 	
For a complete temperature coverage, we examine all AIA channels here.
As shown in \fig{tslice_all.eps} (top), each channel has a generally broad response with one to a few 
peaks. In the order of approximately decreasing temperature response
are 193, 131, 94, 304, 335, 211, and 171~\AA\ channels.
%

\fig{tslice_all.eps} shows space-time plots of all EUV channels.
On the left are base ratio from every other frame 	
of short exposure so that bright flare loops are not over-exposed,
while on the right are running ratio from frames of regular exposure
that is needed for detecting faint emission, including fast loop contractions, above the flare arcade.
  On the left, the 131 and 94~\AA\ channels capture the hot, diffuse loop-top emission,
while the flare arcade underneath it is best seen in cooler channels.
Both regions appear in the 193~\AA\ channel because of 
its dual response to hot \ion{Fe}{24} emission peaking at $18 \MK$
and cool \ion{Fe}{12} emission at $1.6 \MK$ \citep{ODwyer.AIA-T-response.2010A&A...521A..21O}.
  On the right-hand side, downward fast contractions and slow shrinkages can be identified
in the three hottest channels.
 In cooler channels, cooling condensations rain down the arcade loops 	
since 06:00~UT		
at up to $\sim$$170 \kmps$ or 90\% of the free fall speed, faster than typical coronal rain 
\citep{Antolin.coronal-rain-mass-cycle.2012ApJ...745..152A}. 

In \fig{tslice_all.eps} (left), the erupting flux rope is best seen in the hot 131 and 94~\AA\ channels
as a bright, accelerating track,
indicating its high temperature, as also shown in \fig{tem_map.eps}.
Its evolution from gradual rise to impulsive eruption is accompanied by the rapid increase in X-ray flux 
and the loop-top's ascent-to-descent transition.
This is in line with observed synchronous eruption accelerations and flare onsets
\citep{ZhangJie.CME-flare-C1.2001ApJ...559..452Z, ZhangM.flare-timing.2002ApJ...574L..97Z, 
TemmerM.synch.CME-acc.HXR.2008ApJ...673L..95T, LiuW.CaJet2.2011ApJ}.

\figs{lc.eps}(b) and \ref{cent_time.eps}(a) show temporal emission profiles 
taken from the base-ratio space-time plots at $h_{\rm ref}=60 \arcsec$ marked by a cross.
Since 04:17~UT, the 131 and 94~\AA\ emission increases considerably, while other emission
remains essentially flat.
This indicates gentle heating early in the flare, as seen in the 
temperature history at this position shown in \fig{tem_map.eps}(j).
The dip in all channels at the onset of the impulsive phase (vertical dashed line)
is due to the descent of the loop-top source. The large hump afterwards results
from the growth of the flare arcade 	
toward greater heights.		
This hump generally progresses toward cooler channels, indicating a cooling sequence,
similar to those recently observed in flares, AR loops, and prominence condensations
\citep{WoodsT.EVE.flare.2011ApJ...739...59W, Viall.Klimchuk.AR-cooling.2012ApJ...753...35V,
LiuW.Berger.Low.flmt-condense.2012ApJ...745L..21L, Berger.Liu.cavity.condense.2012ApJ...758L..37B}.
The double humps at 193~\AA\ are again due to its dual temperature response. 

Like 193~\AA, the 304~\AA\ channel responds to both hot \ion{Ca}{18} emission
at $7.1~\MK$ and cool \ion{He}{2} emission at $0.05 \MK$.
The dual response of these two channels leads to their unusually large descent
of about 40\% of the loop-top centroids (see \tab{table_cent}).
This is because their maximum loop-top heights are dominated by hot emission,
while the minimum heights are dominated by cool emission at even lower heights
(\fig{cent_time.eps}).

\section*{Appendix B \\
 Disambiguation With \stereo Observations}
\label{append_sect:stereo}

 \begin{figure*}[thbp]      
 \epsscale{1.}
 \plotone{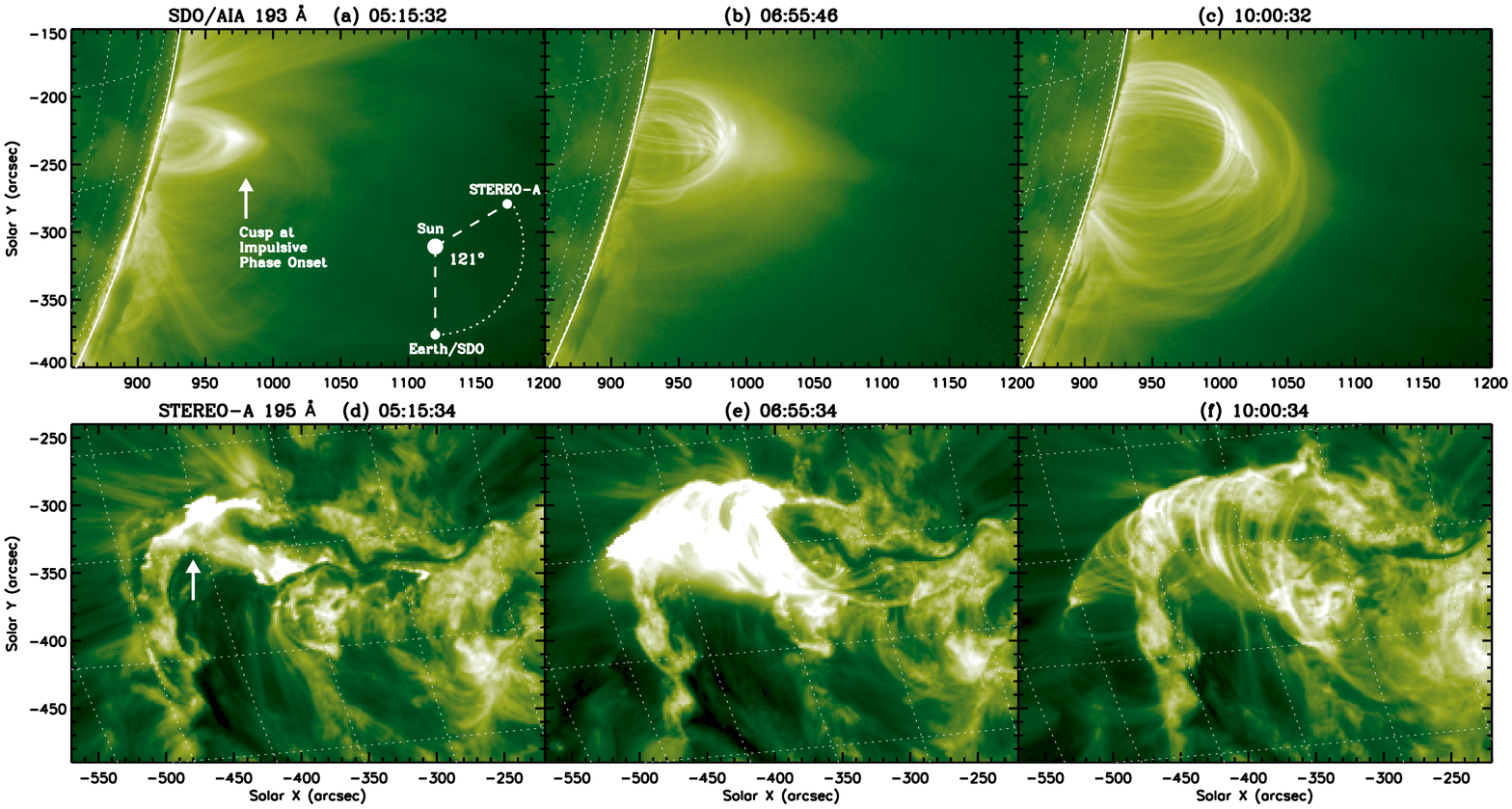}
 \caption[]{\footnotesize
 Comparison of \sdoA/AIA 193~\AA\ (top) and \stereoA-A 195~\AA\ (bottom) images of the flare.
 The \stereo images are scaled as if they are viewed from the same heliocentric distance as \sdoA.
 The heliographic 	
 grids are spaced at $5\degree$.
 Note the compact brightening at the impulsive phase onset (pointed by the arrow),
 in contrast to the extended arcade system later.
 } \label{stereo.eps}
 \end{figure*}
The limb view of \sdoA/AIA leaves an alternative possibility of 
the loop-top descent as the progression of energization
from high- to low-lying loops along the arcade in the east-west direction. 
This possibility is not supported by \stereoA-A (STA) 
that was $121 \degree$ ahead of the Earth and observed the flare on the disk.
When the X-ray loop-top descent occurs, the STA 195~\AA\ image (\fig{stereo.eps}(d))
shows a compact cusp-shaped loop, while the arcade system develops later (panels~(e) and (f)).
One could further argue that a temperature effect, i.e., cooling or heating,
can make a progression appear as a compact source rather than an arcade. 
If so, the loops being cooled or heated would have appeared at the same location 
and been captured by AIA in cooler or hotter EUV channels or by \hsi in X-rays. 
Instead, all available EUV and X-ray data covering a wide temperature range of $10^5$--$10^7 \K$ show
a consistent loop-top descent, which must be a real mass motion.

%



{	\footnotesize	

}




\end{document}